\title{Self-force: Computational Strategies}

\author{Barry Wardell\footnote{Email: barry.wardell@gmail.com, URL: http://www.barrywardell.net} \\
        Department of Astronomy,\\
        Cornell University, Ithaca, NY 14853, USA}

\documentclass[11pt,english]{article}
\usepackage[]{graphicx}
\usepackage[sort&compress,numbers]{natbib}
\usepackage{doi}%<----------
\usepackage{hyperref}
\usepackage{url}
\usepackage{color}
\usepackage[usenames,dvipsnames,svgnames,table]{xcolor}
\usepackage{amsmath}
\usepackage{amssymb}
\usepackage{multirow}

\DeclareMathOperator\erf{erf}

\begin{document}
\maketitle

\begin{abstract}
Building on substantial foundational progress in understanding the effect of a small body's
self-field on its own motion, the past 15 years has seen the emergence of several strategies for
explicitly computing self-field corrections to the equations of motion of a small, point-like
charge. These approaches broadly fall into three categories: (i) mode-sum regularization, (ii)
effective source approaches and (iii) worldline convolution methods. This paper reviews the various
approaches and gives details of how each one is implemented in practice, highlighting some of the
key features in each case.
\end{abstract}

\section{Introduction}
Compact-object binaries are amongst the most compelling sources of gravitational waves. In
particular, the ubiquity of supermassive black holes residing in galactic centres
\cite{Magorrian:1997hw} has made the extreme mass ratio regime a prime target for the eLISA mission
\cite{Gair:2004iv,AmaroSeoane:2007aw,Gair:2008bx,eLISA:GWNotes,eLISA}. Meanwhile, the comparable and 
intermediate mass ratio regimes are an intriguing target for study by the imminent Advanced LIGO 
detector \cite{LIGO}. In order to maximise the scientific gain realised from gravitational-wave
observations, highly accurate models of gravitational-wave sources are essential.
For the case of extreme mass ratio inspirals (EMRIs) --- binary systems in which a compact, solar mass
object inspirals into an approximately million solar mass black hole ---
the demands of gravitational wave astronomy are particularly stringent; the
promise of groundbreaking scientific advances --- including precision tests
of general relativity in the strong-field regime
\cite{Barack:2003fp,Babak:2010ej,Gair:2010bx} and a better census of
black hole populations —-- hinges on our ability to track the phase of their
gravitational waveforms throughout the long inspiral, with an accuracy of better
than 1 part in 10,000 \cite{Gair:2004iv}. This, in turn requires highly
accurate, long time models of the orbital motion.

For the past two decades, these demands have stimulated an intense period of EMRI
research among the gravitational physics community.  Despite of the impressive progress made by
numerical relativists towards tackling the two-body problem in general relativity
(for reviews see Refs.~\cite{Hinder:2010vn,Pfeiffer:2012pc,Sperhake:2014wpa}), the
disparity of length scales characterising the EMRI regime is a significant roadblock
for existing numerical relativity techniques. Indeed, to this day EMRIs remain intractable by current
numerical relativity methods, and successful approaches have instead tackled the problem perturbatively
or through post-Newtonian approximations (see Ref.~\cite{Tiec:2014lba} for a review). This article will focus on the first of these; 
by treating the smaller object as a perturbation to the larger mass, the so-called ``self-force
approach'' reviewed here has been a resoundingly successful tool for EMRI research.

Within the self-force approach, the smaller mass, $\mu<<M$, is assumed to be sufficiently small that it
may be used as a perturbative expansion parameter in the background of the larger mass, $M$.
Expanding the Einstein equation in $\mu$, we see
the smaller object as an effective point particle generating a perturbation about the background of
the larger mass. At zeroth order in $\mu$, the smaller object merely follows a geodesic of the
background. At first order in $\mu$, it deviates from this geodesic due to its interaction with its
self-field. Viewing this deviation as a force acting on the smaller object, the calculation of this
self-force is critical to the accurate modelling of the evolution of the system. For
the purposes of producing an accurate waveform for space-based detectors, and for producing
accurate intermediate mass ratio inspiral (IMRI) models, it will be necessary to include further effects up to
second perturbative order \cite{Isoyama:2012bx,Burko:2013cca}. Indeed, recent compelling
work \cite{LeTiec:2011bk,Tiec:2013twa} suggests that IMRIs —-- and even comparable mass binaries —-- may be modelled using self-force
techniques.

A na\"ive calculation of the first order perturbation due to a point particle leads to a retarded
field which diverges at the location of the particle. The self-force, being the derivative of the
field, also diverges at the location of the particle and one obtains equations of motion which are not
well-defined and must be regularized. A series of formal derivations of the regularized first order
equations of motion (now commonly referred to as the MiSaTaQuWa equations, named after Mino,
Sasaki, Tanaka \cite{Mino:1996nk} and Quinn and Wald \cite{Quinn:1996am} who first derived them)
for a point particle in curved spacetime have been developed
\cite{Dirac:1938nz,DeWitt:1960fc,Hobbs:1968a,Mino:1996nk,Quinn:1996am,Quinn:2000wa,Detweiler:2002mi,
 Harte:2008xq,Harte:2009uy,Harte:2009yr,Galley:2010xn,Galley:2011te}, culminating in a rigorous work by Gralla and Wald
\cite{Gralla:2008fg} and Pound \cite{Pound:2009sm} in the gravitational case and by Gralla, et al.
\cite{Gralla:2009md} in the electromagnetic case. This was subsequently extended to second
perturbative order by Rosenthal \cite{Rosenthal:2005ju,Rosenthal:2005it,Rosenthal:2006nh,Rosenthal:2006iy}, Pound \cite{Pound:2012nt,Pound:2012dk,Pound:2014xva}, Gralla
\cite{Gralla:2012db} and Detweiler \cite{Detweiler:2011tt}. These derivations eliminate the
ambiguities associated with the divergent self-field of a point particle and provide a
well-defined, finite equation of motion. Building upon this foundational progress, several
practical computational strategies have emerged from these formal derivations:
\begin{itemize}
  \item \emph{Dissipative self-force approaches:} While the full first-order self-force
  is divergent, it turns out that the dissipative component is finite and requires no
  regularization. This fact has prompted the development of methods for computing the
  dissipative component alone, sidestepping the issue of regularization altogether.
  These dissipative approaches fall into two categories:
  \begin{enumerate}
    \item \emph{Flux methods:} By measuring the orbit-averaged flux of gravitational waves onto the horizon of the
      larger black hole and out to infinity, the fact that the field is evaluated far away from the worldline
      means that no divergent quantities are ever encountered. This approach yields the time-averaged\footnote{For
      the case of inclined orbits in Kerr spacetime, this is more appropriately formulated as a torus-average.} dissipative
      component of the self-force \cite{Mino:2003yg,Glampedakis:2002ya,Hughes:2005qb,Sago:2005fn,Drasco:2005is,Drasco:2005kz,Sundararajan:2007jg,Fujita:2009us}.
    \item \emph{Local/instantaneous dissipative self-force:}
      The time averaging element of flux methods can be eliminated by instead computing the local, instantaneous
      dissipative component of the self-force from the half-advanced-minus-half-retarded field
      \cite{Mino:2003yg,Gralla:2005et,Hinderer:2008dm,Flanagan:2012kg}.
  \end{enumerate}
  Both methods, however, fundamentally rely on neglecting potentially important
  conservative effects which can significantly alter the orbital phase of the
  system.
  \item The \emph{mode-sum} approach: Introduced in Refs.~\cite{Barack:1999wf,Barack:2001gx}, and
  having since been successfully used in many applications, the approach relies on the decomposition
  of the retarded field into spherical harmonic modes (which are finite, but not differentiable at
  the particle), numerically solving for each mode independently and subtracting
  analytically-derived ``regularization parameters'', then summing over modes.
  \item The \emph{effective source} approach: Proposed in \cite{Barack:2007jh} and
  \cite{Vega:2007mc}, the approach implements the regularization before solving the wave equation.
  This has the advantage that all quantities are finite throughout the calculation and one can directly
  solve a wave equation for the regularized field.
  \item The \emph{worldline convolution} approach: First suggested in
  \cite{Poisson:Wiseman:1998,Anderson:2005gb}, one computes the regularized retarded field as a
  convolution of the retarded Green function along the past worldline of the particle. Although the
  approach is the most closely related to the early formal derivations, it is only recently that it
  has been successfully applied to calculations in black hole spacetimes.
\end{itemize}
For a comprehensive review of the self-force problem, see
Refs.~\cite{Poisson:2011nh,Detweiler:2005kq,Barack:2009ux,Blanchet:2011zz}. In this paper, I will review the various
approaches and give details of how each one is implemented in practice, highlighting the advantages
and disadvantages in each case.

This paper follows the conventions of Misner, Thorne and Wheeler \cite{Misner:1974qy}; a ``mostly
positive'' metric signature, $(-,+,+,+)$, is used for the spacetime metric, the connection
coefficients are defined by
$\Gamma^{\lambda}_{\mu\nu}=\frac{1}{2}g^{\lambda\sigma}(g_{\sigma\mu,\nu}
+g_{\sigma\nu,\mu}-g_{\mu\nu,\sigma}$), the Riemann tensor is
$R^{\alpha}{}_{\!\lambda\mu\nu}=\Gamma^{\alpha}_{\lambda\nu,\mu}
-\Gamma^{\alpha}_{\lambda\mu,\nu}+\Gamma^{\alpha}_{\sigma\mu}\Gamma^{\sigma}_{\lambda\nu}
-\Gamma^{\alpha}_{\sigma\nu}\Gamma^{\sigma}_{\lambda\mu}$, the Ricci tensor and scalar are
$R_{\alpha\beta}=R^{\mu}{}_{\!\alpha\mu\beta}$ and $R=R_{\alpha}{}^{\!\alpha}$, and the Einstein
equations are $G_{\alpha\beta}=R_{\alpha\beta}-\frac{1}{2}g_{\alpha\beta}R=8\pi T_{\alpha\beta}$.
Standard geometrized units are used, with $c=G=1$. Greek indices are used for four-dimensional
spacetime components, symmetrisation of indices is denoted using parenthesis [e.g. $(\alpha
\beta)$], anti-symmetrisation is denoted using square brackets (e.g. $[\alpha \beta]$) and indices
are excluded from symmetrisation by surrounding them by vertical bars [e.g. $(\alpha | \beta |
\gamma)$]. Latin letters starting from $i$ are used for indices summed only over spatial dimensions
and capital letters are used to denote the spinorial/tensorial indices appropriate to the field
being considered. Either $x$ or $x^\mu$ are used when referring to a spacetime field point and
$z(\tau)$ or $z^\mu(\tau)$ are used when referring to a point on a worldline parametrised by proper
time $\tau$. Finally, a retarded (or source) point is denoted using a prime, i.e. $z'$.

\section{Equations of Motion}
The formal equations of motion of a compact object moving in a curved spacetime are now well
established up to second perturbative order. Writing the perturbed spacetime in terms of a
background plus perturbation, $g_{\alpha\beta} = g^{(0)}_{\alpha \beta} + h_{\alpha \beta}$, the
equations of motion essentially amount to those of an accelerated worldline in the background spacetime,
with the acceleration given by a well-defined regular field which is sourced by the worldline.
To order $\mu$, this
coupled system of equations for the worldline and its self-field are commonly referred to as the
MiSaTaQuWa equations and are given (in Lorenz gauge, assuming a Ricci-flat background spacetime) by
\begin{subequations}
\begin{align}
\label{eq:field-gravity}
   \Box \bar{h}^{\rm ret}_{\alpha\beta} + 2 C_{\alpha}{}^{\gamma}{}_{\beta}{}^{\delta } \bar{h}^{\rm ret}_{\gamma\delta} &= - 16 \pi \mu \int g_{\alpha'(\alpha} u^{\alpha'} g_{\beta)\beta'} u^{\beta'} \delta_4(x, z(\tau')) d \tau' \\
\label{eq:accel-gravity}
   \mu \, a^\alpha &= \mu\, k^{\alpha\beta\gamma\delta} \bar{h}^{\rm R}_{\beta\gamma;\delta}
\end{align}
\end{subequations}
with
\begin{equation*}
  k^{\alpha\beta\gamma\delta} = \frac12 g_{(0)}^{\alpha\delta} u^\beta u^\gamma - g_{(0)}^{\alpha\beta} u^\gamma u^\delta - 
  \frac12 u^\alpha u^\beta u^\gamma u^\delta + \frac14 u^\alpha g_{(0)}^{\beta \gamma} u^\delta + 
 \frac14 g_{(0)}^{\alpha \delta} g_{(0)}^{\beta \gamma}.
\end{equation*}
Here, $\mu$ is mass of the object, $g_{\alpha'\alpha}$ is the bivector of parallel transport,
$C_{\alpha\beta\gamma\delta}$ is the Weyl tensor of the background
spacetime, and we use the trace-reversed metric perturbation
$\bar{h}_{\alpha\beta} = h_{\alpha\beta} - \frac{1}{2} g^{(0)}_{\alpha\beta} h$, $h = h^\gamma{}_{\gamma}$. 

One can also consider compact objects possessing other types of charge. For example,
a particularly simple case is that of a scalar charge $q$ with mass $m$ and scalar field $\Phi$,
in which case the equations of motion are given by\footnote{In the scalar case, it is important to distinguish
between the self-force $F^\alpha = \nabla^\alpha \Phi$ and the self-acceleration, which is given
by projecting to self-force orthogonal to the worldline,
$a^\alpha = (g_{(0)}^{\alpha \beta} + u^{\alpha} u^{\beta}) F_{\beta}$. In the electromagnetic and
gravitational cases the self-force has no component along the worldline and the two may be used
interchangeably.}${}^{,}$\footnote{We assume that the mass $m$ is small and ignore its effect on the equations
of motion.}
\begin{subequations}
\begin{align}
\label{eq:field-scalar}
  (\Box - \xi R) \Phi^{\rm ret} &= - 4 \pi q \int \delta_4(x, z(\tau')) d \tau' \\
\label{eq:accel-scalar}
  m a^\alpha &= q \big(g_{(0)}^{\alpha\beta}+u^\alpha u^\beta\big) \Phi^{\rm R}_{,\beta} \\
\label{eq:mdot-scalar}
  \frac{dm}{d\tau} &= - q \, u^\alpha \Phi^{\rm R}_{,\alpha}.
\end{align}
\end{subequations}
Here, $R$ is the Ricci scalar of the background spacetime and $\xi$ is the coupling to scalar curvature.
Similarly, for an electric charge, $e$, one obtains equations of motion which
are given in Lorenz gauge by
\begin{subequations}
\begin{align}
\label{eq:field-em}
  \Box A^{\rm ret}_\alpha - R_\alpha^\beta A^{\rm ret}_\beta &= - 4 \pi e \int g_{\alpha\alpha'} u^{\alpha'} \delta_4(x, z(\tau')) d \tau' \\
\label{eq:accel-em}
  m a^\alpha &= e \big(g_{(0)}^{\alpha\beta}+u^\alpha u^\beta\big) A^{\rm R}_{[\gamma, \beta]} u^\gamma,
\end{align}
\end{subequations}
where $R_{\alpha\beta}$ is the Ricci tensor of the background spacetime and $A^\mu$ is the vector potential.

The key component in all instances
is the identification of the appropriate regularized field on the worldline. Detweiler and Whiting
identified a particularly elegant choice for the regularized field, written in terms of the
difference between the retarded field and a locally-defined singular field,
\begin{equation}
\label{eq:singular-regular-split}
  \Phi^{\rm R} = \Phi^{\rm ret} - \Phi^{\rm S}, \quad
  A^{\rm R}_\alpha = A^{\rm ret}_\alpha - A^{\rm S}_\alpha, \quad
  h^{\rm R}_{\alpha\beta} = h^{\rm ret}_{\alpha\beta} - h^{\rm S}_{\alpha\beta}.
\end{equation}
In addition to giving the physically-correct self-force, the Detweiler-Whiting regular field has the
appealing feature of being a solution of the homogeneous field equations in the vicinity of the
worldline. Most computational strategies essentially amount to differing ways of representing this
singular field\footnote{Some methods \cite{Lousto:2008mb,Kol:2013tfa} rely on alternative
prescriptions for the singular field than that proposed by Detweiler and Whiting.}
and obtaining the regularized field on the worldline.

\section{Numerical regularization strategies}
In a numerical implementation, it is essential to avoid the evaluation of divergent quantities. In
the case of self-force calculations, both the retarded and singular fields diverge on the worldline
so one must avoid evaluating them there. Several strategies for doing so have emerged over the
years (see Table \ref{table:methods} for a summary).

One option is to only ever evaluate finite, dissipative quantities (e.g. the retarded
field far from the worldline or the half-advanced-minus-half-retarded field on
the worldline).
This is the basis of the dissipative methods mentioned in the
introduction. Since these methods effectively avoid the problem of
regularization, they will not be discussed further here; we will return to them
in Sec.~\ref{sec:evolution}. It is worth noting, however, that the methods
typically used by dissipative calculations are essentially the same as those
used by mode-sum regularization for computing the retarded field, but without
the additional regularization step.

This leaves three regularization strategies which allow the regularized field to be computed on the
worldline without encountering numerical divergences: worldline convolution, mode-sum
regularization, and the effective source approach.

\begin{table}
\begin{tabular}{|c|m{2mm}||m{3.5cm}|m{3.5cm}|m{3.5cm}|}
\hline
 \multicolumn{2}{|c||}{Case} & Worldline & Mode-sum & Effective Source \\
\hline
\parbox[t]{2mm}{\multirow{2}{*}{\rotatebox[origin=c]{90}{Scalar\qquad \,}}}
&
\rotatebox[origin=c]{90}{\,Schwarzschild\,}
&
circular (apprx)~\cite{Anderson:2005gb};\hfill\break
generic\hfill\break(quasilocal)
\cite{Casals:2009xa,Ottewill:2007mz};\hfill\break
generic~\cite{Casals:2013mpa,Wardell:2014kea,Zenginoglu:2012xe};\hfill\break
static~\cite{Casals:2012qq};\hfill\break
accelerated~\cite{Ottewill:2008uu};
&
radial~\cite{Barack:2000zq};\hfill\break
circular~\cite{Burko:2000xx,Detweiler:2002gi,DiazRivera:2004ik,Canizares:2009ay};\hfill\break
eccentric~\cite{Haas:2007kz,Haas:2006ne,Canizares:2010yx,Heffernan:2012su,Thornburg:2010tq};\hfill\break
static~\cite{Casals:2012qq};
&
circular~\cite{Barack:2007jh,Vega:2007mc,Dolan:2010mt,Lousto:2008mb,Vega:2009qb,Warburton:2013lea};\hfill\break
eccentric~\cite{Vega:2013wxa};\hfill\break
evolving~\cite{Diener:2011cc};
\\
\cline{2-5}
&
\rotatebox[origin=c]{90}{Kerr}
&
generic~\cite{Ottewill:2007mz};\hfill\break
accelerated~\cite{Ottewill:2008uu};
&
circular~\cite{Warburton:2010eq};\hfill\break
equatorial~\cite{Heffernan:2012vj,Warburton:2011hp};\hfill\break
inclined circular~\cite{Warburton:2014bya};\hfill\break
accelerated~\cite{Linz:2014pka};\hfill\break
static~\cite{Burko:2001kr,Ottewill:2012aj};
&
circular~\cite{Dolan:2011dx};\hfill\break
eccentric~\cite{Thornburg:Capra17};
\\
\hline
\parbox[t]{2mm}{\multirow{2}{*}{\rotatebox[origin=c]{90}{EM\qquad\,}}}
&
\rotatebox[origin=c]{90}{\,Schwarzschild\,}
&
static~\cite{Casals:2012qq};
&
static~\cite{Casals:2012qq};\hfill\break
eccentric~\cite{Haas:2011np,Heffernan:2012su};\hfill\break
static~(Schwarzschild-de~Sitter)~\cite{Kuchar:2013bla};\hfill\break
radial~(Reissner-\hfill\break
Nordstr\"om)~\cite{Zimmerman:2012zu};
&
---
\\
\cline{2-5}
&
\rotatebox[origin=c]{90}{\,Kerr\,}
&
---
&
equatorial~\cite{Heffernan:2012vj};\hfill\break
accelerated~\cite{Linz:2014pka};
&
---
\\
\hline
\parbox[t]{2mm}{\multirow{2}{*}{\rotatebox[origin=c]{90}{Gravity\quad\,}}} & \rotatebox[origin=c]{90}{\,Schwarzschild\,}
&
generic\hfill\break
(quasilocal)~\cite{Anderson:2004eg};
&
radial~\cite{Barack:2002ku};\hfill\break
circular~\cite{Barack:2005nr,Dolan:2013roa,Dolan:2014pja,Field:2010xn,Keidl:2010pm,Merlin:2014qda,Sago:2008id,Shah:2010bi,Akcay:2010dx};\hfill\break
eccentric~\cite{Heffernan:2012su,Barack:2008ms,Barack:2010tm,Sago:2009zz,Field:2009kk,Hopper:2010uv,Hopper:2012ty,Akcay:2013wfa,Pound:2013faa,Osburn:2014hoa};\hfill\break
osculating~\cite{Warburton:2011fk};
&
circular~\cite{Dolan:2012jg};
\\
\cline{2-5}
&
\rotatebox[origin=c]{90}{\,Kerr\,}
&
circular\hfill\break
(quasilocal)~\cite{Anderson:2005gb};\hfill\break
branch cut~\cite{Kavanagh:Capra17};
&
equatorial~\cite{Heffernan:2012vj};\hfill\break
accelerated~\cite{Linz:2014pka};\hfill\break
circular~\cite{Pound:2013faa,Shah:2012gu};
&
circular~\cite{Dolan:Capra16};\hfill\break
generic~\cite{Wardell:2011gb};
\\
\hline
\end{tabular}
\caption{Summary of regularization methods employed by self-force calculations in black hole spacetimes.}
\label{table:methods}
\end{table}

\subsection{Worldline convolution}
\begin{figure}[htb!]
\begin{center}
\includegraphics[width=6cm]{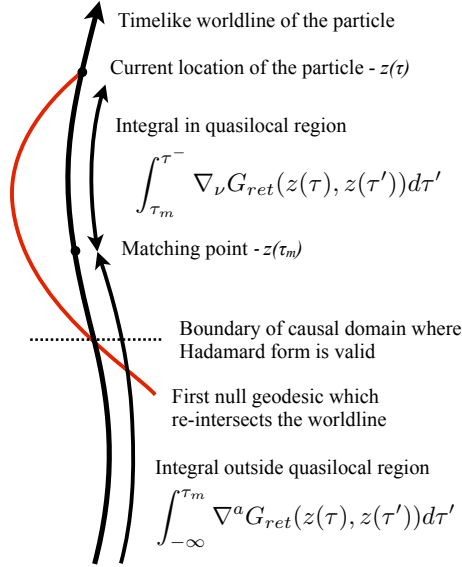}
\end{center}
\caption{
Schematic representation of the worldline convolution method. The equations shown are for the case
of the self-force on a scalar charge, but are representative of similar equations in the
electromagnetic and gravitational cases. Reproduced from Ref.~\cite{Casals:2013mpa}.
}
\label{fig:convolution}
\end{figure}

The worldline convolution method relies on a split of the regularized self-force into an
``instantaneous'' piece and a history-dependent term. The instantaneous piece is easily calculated
from local quantities evaluated at the particle's position,
\begin{subequations}
\label{eq:inst}
\begin{align}
\Phi_{,\alpha}^{\text{inst}} &= q \bigg[\frac{1}{2}\Big(\xi - \frac{1}{6}\Big)R u_\alpha + \Big(g^{(0)}_{\alpha\beta} + u_\alpha u_\beta\Big) \Big(\frac{1}{3} \dot{a}^\beta + \frac{1}{6} R^{\beta}{}_{\gamma} u^\gamma\Big)\bigg], \\
A_{[\beta;\alpha]}^{\text{inst}} u^\beta &= e \Big(g^{(0)}_{\alpha\beta} + u_\alpha u_\beta\Big) \Big(\frac{1}{3} \dot{a}^\beta + \frac{1}{6} R^{\beta}{}_{\gamma} u^\gamma\Big), \\
\bar{h}_{\alpha\beta;\gamma}^{\text{inst}} u^\beta u^\gamma & = \bar{h}_{\alpha\beta;\gamma}^{\text{inst}} u^\alpha u^\beta = 0.
\end{align}
\end{subequations}
The history-dependent term is much more difficult to calculate, as it is given in terms of
a convolution of the derivative of the retarded Green function along the worldline's entire
past-history (see Fig.~\ref{fig:convolution}),
\begin{subequations}
\label{eq:hist}
\begin{align}
\Phi_{,\alpha}^{\text{hist}} &= q \int_{-\infty}^{\tau^-} \nabla_\alpha G^{\rm ret}[x, z(\tau')] \,d\tau',\\
A_{\alpha;\beta}^{\text{hist}} &= e \int_{-\infty}^{\tau^-} \nabla_\beta G^{\rm ret}_{\alpha\alpha'}[x, z(\tau')] u^{\alpha'}\,d\tau',\\
\bar{h}_{\alpha\beta;\gamma}^{\text{hist}} &= 4\mu \int_{-\infty}^{\tau^-} \nabla_\gamma G^{\rm ret}_{\alpha\beta\alpha'\beta'}[x, z(\tau')] u^{\alpha'} u^{\beta'} \,d\tau'.
\end{align}
\end{subequations}
The covariant derivatives here are taken with respect to the first argument of the retarded Green
function. Likewise, the regularized self-field can be obtained from a worldline convolution of the
retarded Green function itself; similar formulae can also be derived for higher derivatives.
The retarded Green function appearing in these equations is a solution of the wave equation with a
delta-function source,
\begin{subequations}
\label{eq:BoxG}
\begin{align}
\label{eq:BoxG-scalar}
(\Box - \xi R) G^{\rm ret} &= - 4 \pi \, \delta^4 (x,x'), \\
\Box G^{\rm ret}_{\alpha\alpha'}
  - R_\alpha{}^\beta G^{\rm ret}_{\beta\alpha'} &=
  - 4 \pi \, g_{\alpha\alpha'} \delta^4 (x,x'), \\
\Box G^{\rm ret}_{\alpha \beta\alpha' \beta'}
  + 2 R_\alpha{}^\gamma{}_\beta{}^\delta G^{\rm ret}_{\gamma \delta\alpha' \beta'} &=
  - 4 \pi \, g_{\alpha\alpha'} g_{\beta\beta'} \delta^4 (x,x'),
\end{align}
\end{subequations}
with boundary conditions such that the solutions correspond to purely outgoing radiation at infinity and no radiation emerging from the horizon.

The regularization (i.e. subtraction of the Detweiler-Whiting singular field) is formally achieved
by the limiting procedure in the upper limit of integration, i.e. by cutting off the integration at
$\tau^-$, slightly before the coincidence point $z = x$. In practice, this is done by taking the
integration all the way up to coincidence, but excluding the \emph{direct} contribution to the
Green function at coincidence (i.e. the term proportional to $\delta(\sigma)$ in the Hadamard form for
the Green function). Although the retarded Green function also diverges at certain other points
along the past worldline (in particular at null-geodesic intersections where the particle sees null
rays emitted from its past), it turns out that these are all integrable singularities (of the form
$1/\sigma$ and $\delta(\sigma)$) and the integral may be accurately evaluated to give a finite and
accurate value for the regularized self-force.

Despite having been proposed in the early days of EMRI self-force calculations
\cite{Poisson:Wiseman:1998}, the worldline convolution method was largely ignored for a long time by
numerical implementations (the notable exception being investigative studies
\cite{Anderson:2005gb,Capon:1998} which did not complete a full calculation of the
self-force). A likely reason is that the method relies on knowledge of the retarded Green function
for all points on the particle's past worldline. While methods have existed for decades for 
computing portions of the retarded Green function, it turns out to be very difficult to obtain it
accurately everywhere that it is needed for the worldline convolution.

Thankfully, recent progress has led to two practical computational strategies, both of which have
been successfully applied to compute the self-force in black hole spacetimes. The first of these is
based on a frequency-domain decomposition of the Green function, and builds on the rich history of
black hole perturbation theory developed over the past several decades. The second, a time-domain
approach, has been a very recent development and shows a great deal of promise for the future.

Independently of whether a frequency-domain or a time-domain scheme is used, a common problem with
both is that they fail at early times when source and field points are close together; in the
frequency-domain case the convergence is poor at early times, while in the time-domain case
features from the numerical approximation pollute the data at early times. A relatively
straightforward solution which has been successfully applied in both scenarios is to only rely on
their results at late times and to supplement them at early times with a quasilocal Taylor series
expansion. Provided a sufficiently early time can be chosen where both the distant past and
quasilocal calculations converge, this yields a global approximation for the Green function which
is sufficient for producing an accurate result for the self-force. To date this has been shown to
be possible in the case of a scalar charge in Nariai \cite{Casals:2009zh},
Schwarzschild \cite{Casals:2013mpa,Wardell:2014kea} and Kerr \cite{Wardell:Capra17} spacetimes.

Given an approximation for the Green function valid throughout the past worldline, it is then
trivial to numerically integrate Eq.~\eqref{eq:hist} to obtain the self-force
(see Fig.~\ref{fig:convolution}). As mentioned previously, the divergences in the retarded Green function at null-geodesic intersections on the past worldline are all integrable singularities and do not pose a significant obstacle to accurate numerical evaluation of the integrals.

\subsubsection{Quasilocal expansion}

In order to obtain an approximation to the retarded Green function which is valid at early times,
it is convenient to start with the Hadamard form for the Green function (in Lorenz gauge)
\begin{equation}
G^{\mathrm{ret}}_{AB'}(x,x') = \Theta_{-} \Big[ U_{AB'} \delta (\sigma) - V_{AB'} \Theta (-\sigma) \Big] ,
\end{equation}
where $\Theta_{-}$ is analogous to the Heaviside step-function, being $1$ when $x'$ is in the
causal past of $x$, and $0$ otherwise, $\delta(\sigma)$ is the covariant form of the Dirac delta
function, $U_{AB'}$ and $V_{AB'}$ are symmetric bi-spinors/tensors and are regular for
$x'\rightarrow x$. The bi-scalar $\sigma \left(x,x'\right)$ is the Synge world function, which is
equal to one half of the squared geodesic distance between $x$ and $x'$. In particular,
$\sigma(x,x) = 0$ and $\sigma(x,z)<0$ when $x$ and $z$ are timelike separated. Because of the
limiting procedure in the history integral, Eq.~\eqref{eq:hist}, only the term involving $V_{AB'}$
is non-zero, and at all required points along the worldline $\Theta_- = 1 = \Theta(-\sigma)$. The
problem of determining the retarded Green function at early times therefore reduces to finding an
approximation for $V_{AB'}(x,x')$ which is valid for $x$ and $x'$ close together.

Several methods have been developed for computing approximations to $V_{AB'}$. Fundamentally, they
rely on either the use of a series expansion, or on the use of numerically evolved transport
equations (ordinary differential equations defined along a worldline). The series expansion
approach has been the most fruitful to date with results including: leading-order coordinate
expansions in Schwarzschild and Kerr spacetimes for scalar \cite{Ottewill:2007mz,Ottewill:2008uu}
and gravitational cases \cite{Anderson:2004eg}, high-order coordinate expansions in spherically
symmetric spacetimes (including Schwarzschild) \cite{Casals:2009xa}, formal covariant expansions
in generic spacetimes \cite{Avramidi:1986mj,Avramidi:2000pia,Decanini:2005gt,Ottewill:2009uj},
and moderately high-order coordinate expansions in Schwarzschild \cite{Heffernan:2012su} and Kerr
\cite{Heffernan:2012vj} spacetimes. The only numerical calculation I am aware of was done in
\cite{Ottewill:2009uj} for generic spacetimes (with an example application in Schwarzschild
spacetime).

The series expansion method produces an expression for $V(x,x')$ as a power series in the coordinate
distance between $x$ and $x'$. For example, for the scalar case in Schwarzschild spacetime it
takes the form
\begin{equation} \label{eq:CoordGreen}
V(x,x') =
 \sum_{i,j,k=0}^{\infty} v_{ijk}(r) ~ (t-t')^{2i} (1-\cos\gamma)^j (r-r')^k,
\end{equation}
where $\gamma$ is the angular separation of the points and the $v_{ijk}$ are analytic functions of
$r$ and $M$. It is straightforward to take partial derivatives of these expressions at either
spacetime point to obtain the derivative of the Green function. Although this series on its own may
be sufficient for use in the quasilocal component of a worldline convolution, it turns out that
some simple tricks allow for a vast improvement in accuracy. It turns out that, because
$V(x,x')$ diverges at the edge of the normal neighbourhood, the series approximation benefits
significantly from Pad\'e resummation which incorporates information about the form of the
divergence \cite{Casals:2009xa}. One minor caveat is that since Pad\'e re-summation is only well defined for
series expansions in a single variable, it is necessary to first expand $r'$ and $\gamma$ in a
Taylor series in $t-t'$, using the equations of motion to determine the higher derivatives
appearing in the series coefficients. Then, with $V(x,x')$ written as a power series in $t-t'$
alone, a standard diagonal Pad\'e approximant provides an accurate representation of the Green
function in the quasi-local region.

\subsubsection{Frequency domain methods}
\label{sec:GF-FD}
Frequency domain methods for computing the retarded Green function rely crucially on the
separability of the wave equation. In the scalar, Schwarzschild case that represents the current
state-of-the art\footnote{More generally, Teukolsky \cite{Teukolsky:1972my,Teukolsky:1973ha} showed that the field
equations may be separated in Kerr spacetime in the gravitational case by making use of the
spin-weighted spheroidal harmonics in place of the spherical harmonics. A series of works --- pioneered
by Regge and Wheeler \cite{Regge:1957td} and improved upon by others
\cite{Zerilli:1971wd,Vishveshwara:1970cc,Moncrief:1974am,Cunningham:1978zfa,Cunningham:1979px,Martel:2005ir,Berndtson:2009hp} --- achieved a
similar separation in the Schwarzschild gravitational case by making use of tensor spherical
harmonics. However, separability alone is not sufficient and there remain some technical issues
which have yet to be solved before solutions of the Teukolsky or Regge-Wheeler equations could be
used in a worldline convolution approach. Most important is the issue of gauge; the MiSaTaQuWa
equations, Eqs.~\eqref{eq:inst} and \eqref{eq:hist}, were derived in the Lorenz gauge, whereas
solutions of the Teukolsky and Regge-Wheeler equations are in a gauge different from Lorenz gauge.
Fortunately, recent work \cite{Pound:2013faa} has resolved many of the conceptual issues associated
with gauge choice.}
\cite{Casals:2013mpa} (also see \cite{Casals:2009zh} for a related calculation in Nariai
spacetime), this can be achieved by writing the Green function as a sum of spherical harmonic and
Fourier modes
\begin{equation}
\label{eq:G-FD}
G^{\rm ret}(x,x')= \frac{1}{r\, r'} \sum_{\ell=0}^{\infty}
 P_{\ell}(\cos\gamma)\frac{1}{2\pi}
\int_{-\infty+i \epsilon}^{\infty+i \epsilon} \hat{g}_{\ell}(r,r';\omega) e^{-i\omega (t-t')} d\omega.
\end{equation}
Here, $\epsilon>0$ is a formal parameter to ensure the correct boundary conditions are satisfied
for a retarded Green function. Substituting this into the wave equation,
Eq.~\eqref{eq:BoxG-scalar}, one obtains an independent set of ordinary differential equations for
$\hat{g}_\ell(r,r'; \omega)$, one equation for each $\ell$ and $\omega$,
\begin{equation}
\label{eq:radial}
\left[\frac{d^2}{dr_*^2}+\omega^2-V_\ell(r)\right]\hat{g}_{\ell}(r,r';\omega)=-\delta(r_*-r'_*),
\end{equation}
with
\begin{equation}
  V_\ell(r) \equiv \left(1-\frac{2M}{r}\right)\left[\frac{\ell(\ell+1)}{r^2}+\frac{2M}{r^3}\right].
\end{equation}
Here, $r_*\equiv r+2M\ln\left(\frac{r}{2M}-1\right)$ is the radial tortoise coordinate.

Given two linearly-independent solutions, $p(r;\omega)$ and $q(r;\omega)$, of the homogeneous
version of (\ref{eq:radial}), $\hat{g}_\ell(r,r';\omega)$ is given by
\begin{align}
\hat{g}_\ell(r,r';\omega)=\frac{p_\ell(r_<,\omega)q_\ell(r_>,\omega)}{W(p,q)},
\end{align}
where $r_>\equiv \max(r,r'),\ r_<\equiv \min(r,r')$ and $W(p,q)$ is the Wronskian. One is then
faced with the integral over frequencies in the inverse Fourier transform appearing in
\eqref{eq:G-FD}.
\begin{figure}[htb!]
\begin{center}
\includegraphics[width=6.1cm]{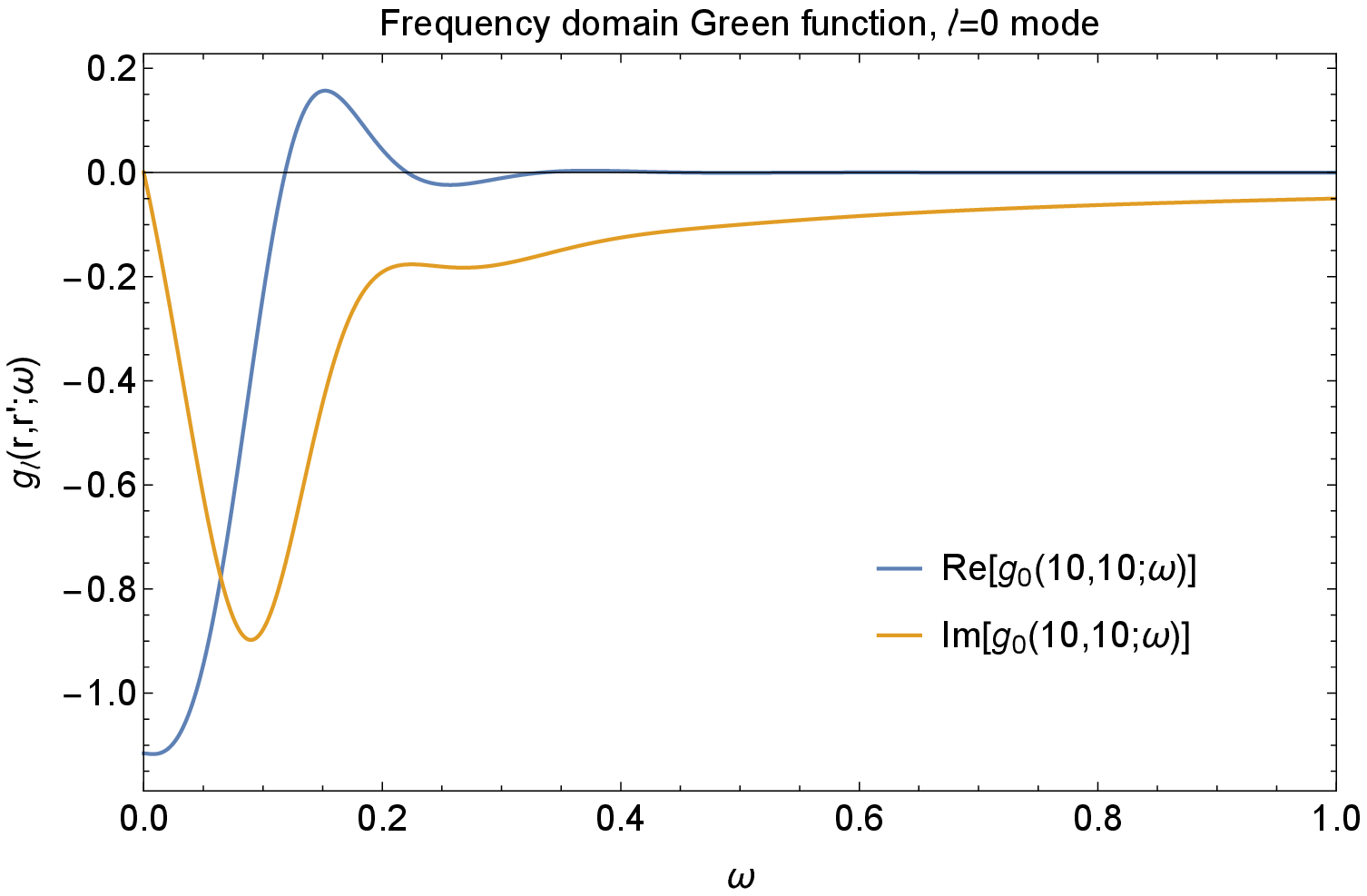}
\includegraphics[width=6.1cm]{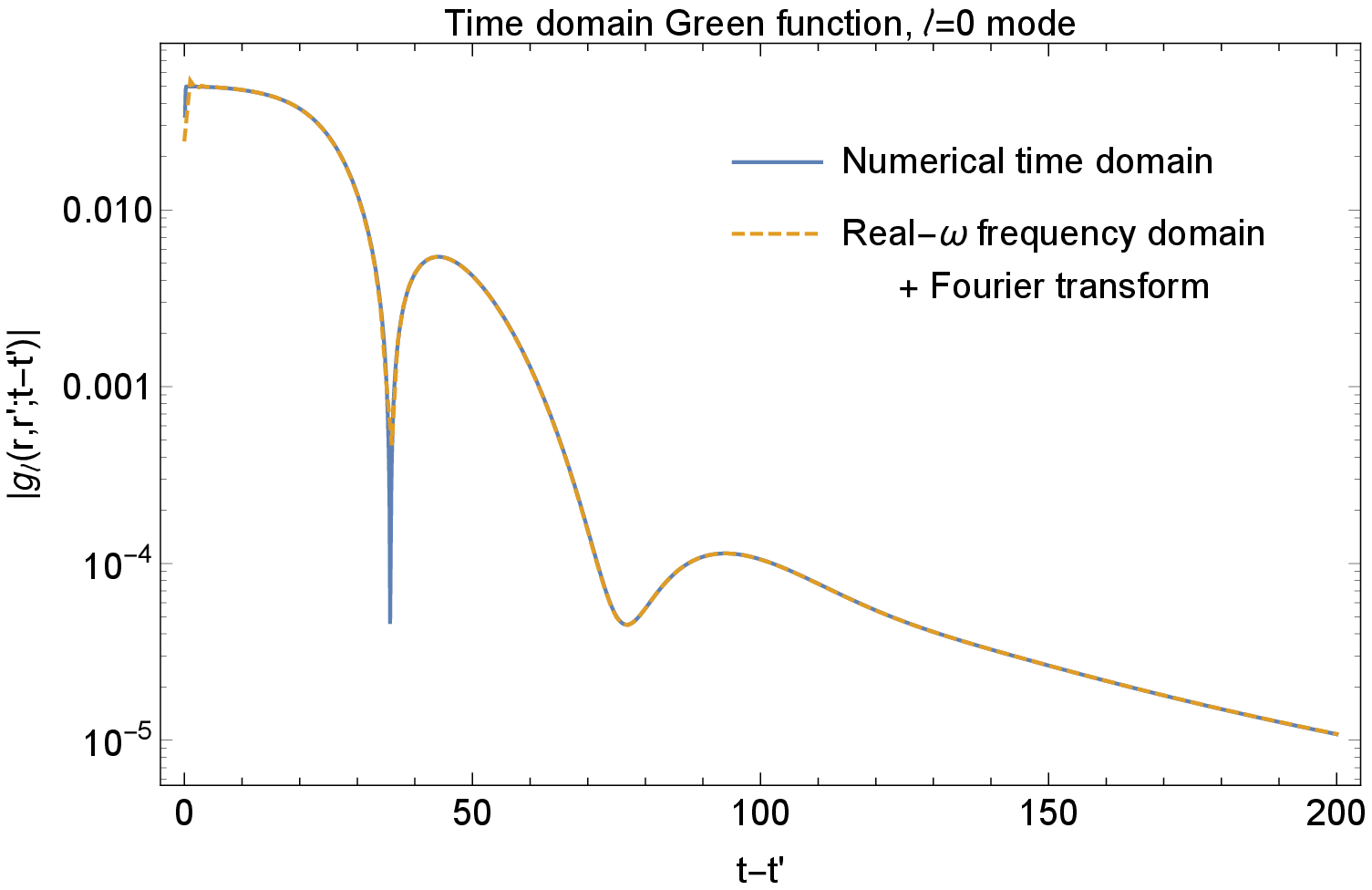}
\includegraphics[width=6.1cm]{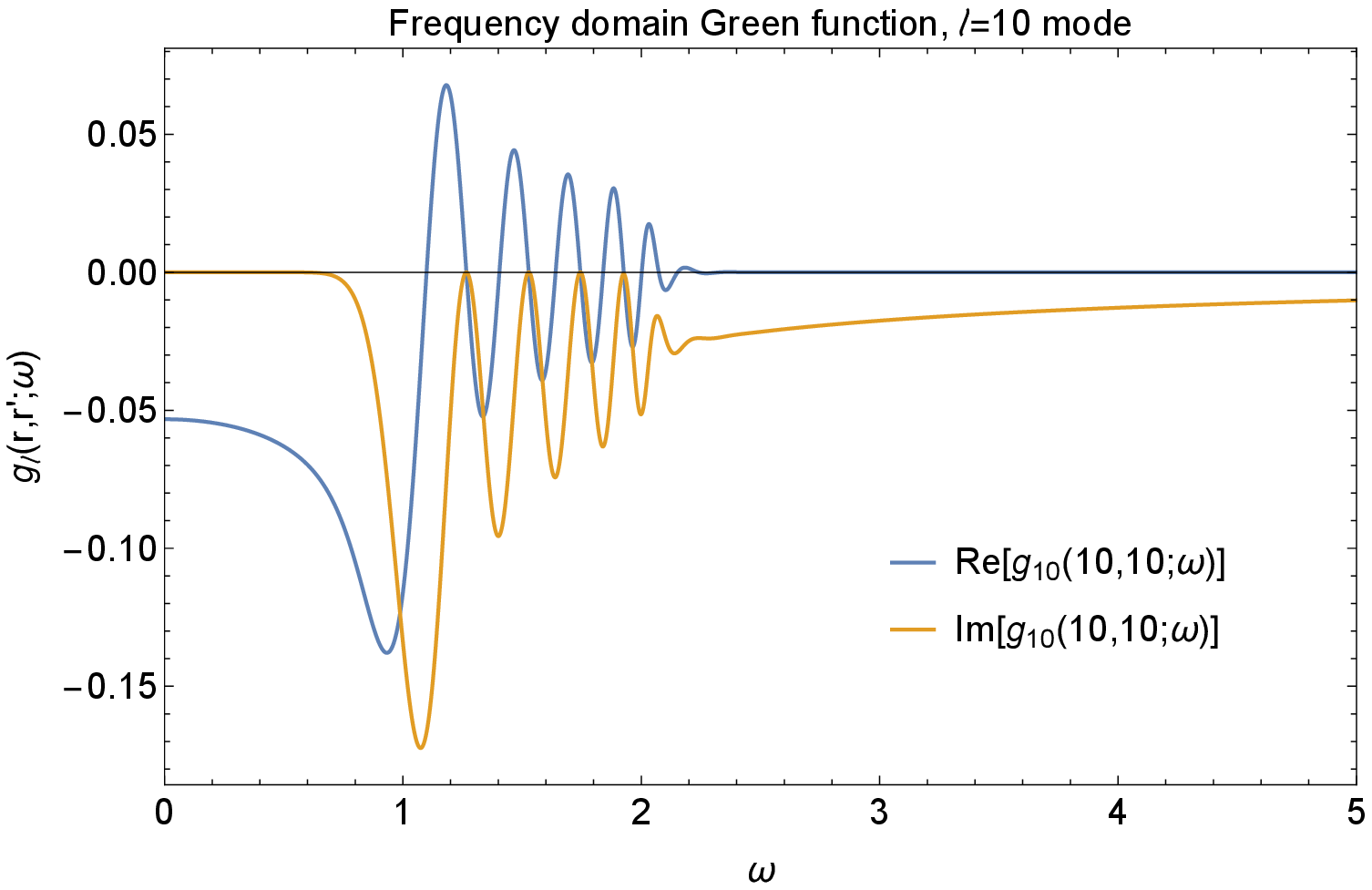}
\includegraphics[width=6.1cm]{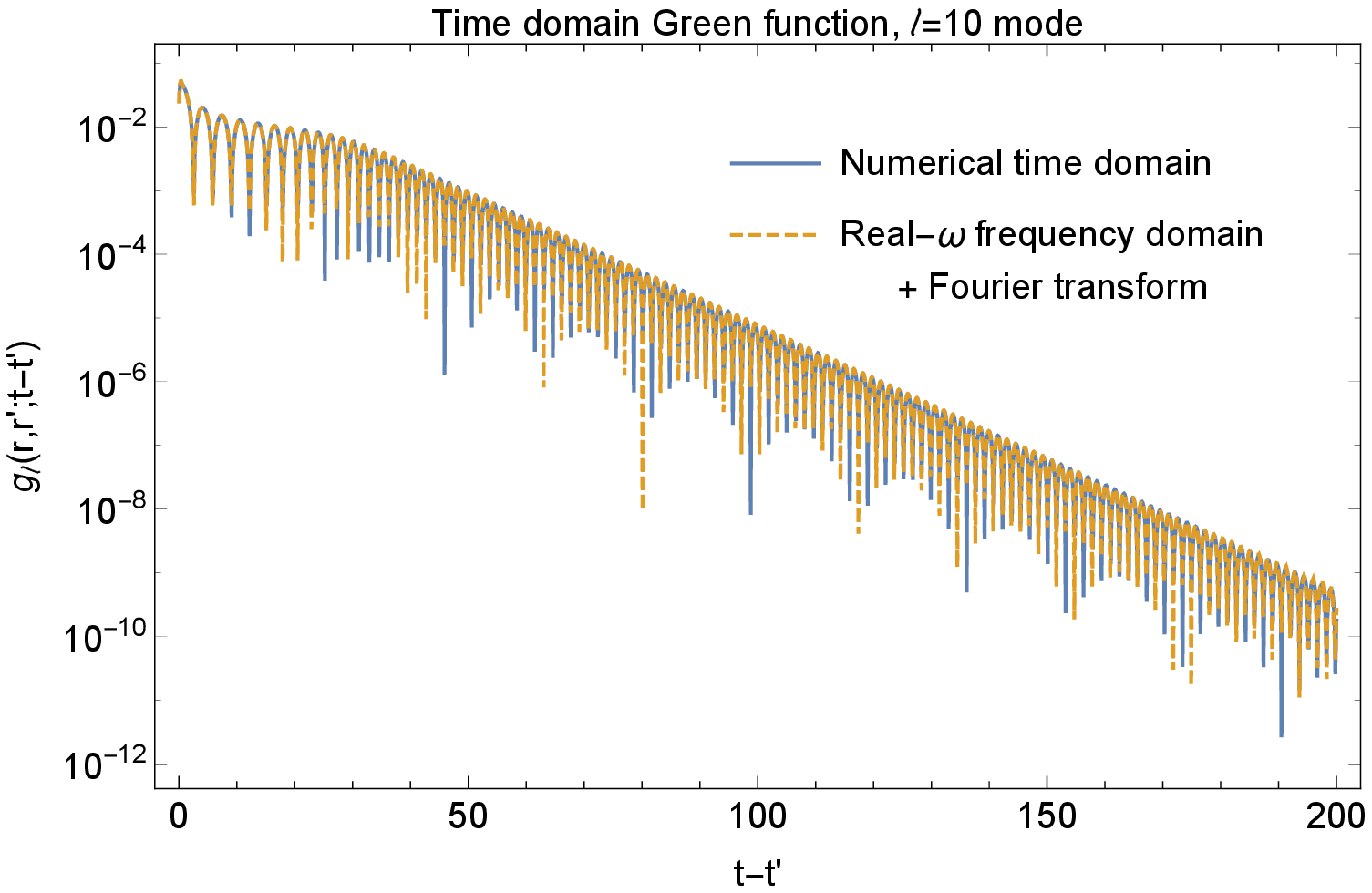}
\end{center}
\caption{
Spherical harmonic modes of the Schwarzschild scalar Green function. Left: the frequency domain
Green function, $\hat{g}_\ell (r,r'; \omega)$, as a function of real frequency for $r=r'=10M$, $\ell = 0$
(top) and $\ell=10$ (bottom). Right: the corresponding time domain Green function, $g_\ell(r,r';
t-t')$ computed by Fourier transforming the frequency domain Green function using
Eq.~\eqref{eq:GF-fourier} with $\omega_{\rm max} = 8.5$ (blue, solid line), and by using the time
domain methods described in Sec.~\ref{sec:GF-TD} (orange, dashed line).
}
\label{fig:GF-ell}
\end{figure}
This may be achieved through straightforward integration along the real-$\omega$
axis (see Fig.~\ref{fig:GF-ell}); the only caveat is that the formally-infinite integral over frequencies should be cut off at some finite maximum frequency using a smooth window function, for example
\begin{equation}
  \label{eq:GF-fourier}
  g_{\ell} (r, r'; t-t')  = \frac{1}{2\pi} \int_{-\infty+i \epsilon}^{\infty+i \epsilon} \hat{g}_{\ell}(r,r';\omega) e^{-i\omega (t-t')} \big(1-\erf[2(\omega - \omega_{\rm max})]\big)/2 \, d\omega.
\end{equation}

\begin{figure}[htb!]
\begin{center}
\includegraphics[width=7cm]{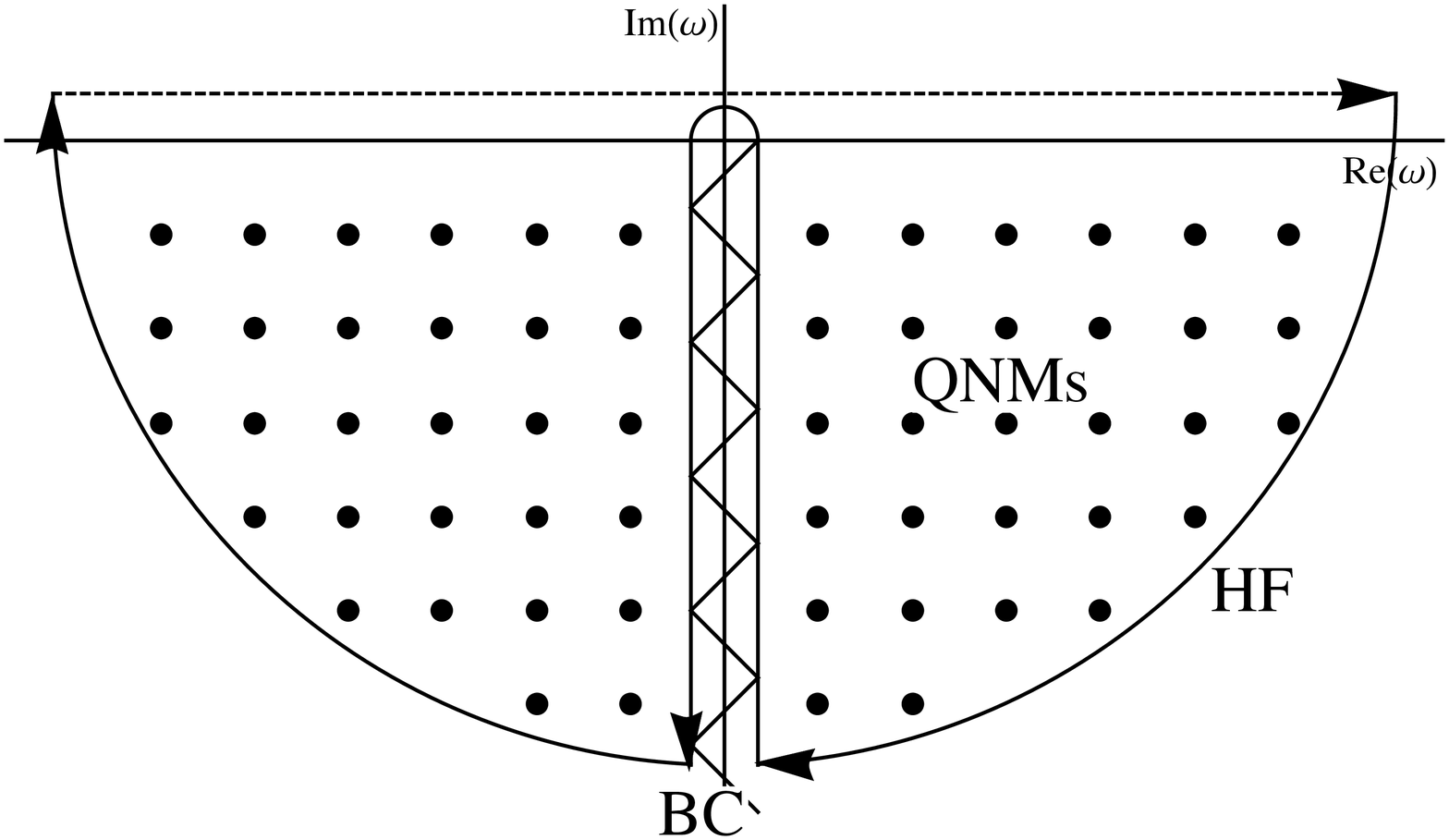}
\end{center}
\caption{
Contour deformation in the complex-frequency plane. The residue theorem of complex analysis allows
one to re-express the integral over (just above) the real line of the Fourier modes,
$\hat{g}_\ell(r,r';\omega)$, as an integral over a high-frequency arc plus an integral around a branch cut
and a sum over the residues at the poles.
Reproduced from Ref.~\cite{Casals:2013mpa}.
}
\label{fig:contour}
\end{figure}

Alternatively, as was done in \cite{Casals:2013mpa}, the integration contour can be deformed into
the complex frequency plane following a proposal by Leaver \cite{Leaver:1986gd,PhysRevD.38.725}. In
Schwarzschild spacetime,
$\hat{g}_\ell(r,r';\omega)$ has simple poles in the lower semi-plane at the quasinormal mode frequencies,
along with a branch cut starting at $\omega=0$ and continuing along the negative imaginary axis
(see Fig.~\ref{fig:contour}). The deformed frequency integral is therefore given by an integral
over a high-frequency arc, an integral around a branch cut, and an integral over the residues at
the poles\footnote{In the Kerr spacetime there are indications that there may be additional branch
cuts to consider \cite{Kavanagh:Capra17}.}.

The high-frequency arcs can be disregarded as they are likely to only contribute at very early times,
where the quasilocal approximation can be used. There are well-established methods for accurately
computing the location of the poles (quasinormal mode frequencies) and the residues at the poles.
The biggest difficulty in the frequency-domain approach is in evaluating the branch cut integral,
about which very little was known until recently (other than asymptotic approximations for e.g.
large radius or late times). Substantial recent progress has established methods for calculating
this branch cut contribution \cite{Casals:2011aa}; these methods were used in \cite{Casals:2013mpa} to compute the self-force in Schwarzschild spacetime.

The final step in the frequency domain approach is to sum over spherical harmonic modes to produce
the full Green function. Here, the distributional parts of the full Green function can cause poor
convergence in the mode-sum. Fortunately, there is a straightforward solution to this problem: by
smoothly cutting off the mode sum at large $\ell$,
\begin{equation}
G^{\rm ret}(x,x')= \frac{1}{r\, r'} \sum_{\ell=0}^{\ell_{\rm max}}
 P_{\ell}(\cos\gamma) g_{\ell}(r,r'; t-t') e^{-\ell^2 / 2 \ell_{\rm cut}^2}
\end{equation}
one obtains a mollified retarded Green which is appropriate for use in a self-force calculation,
and whose sum over $\ell$ converges. Empirically, it has been found that choosing
$\ell_{\rm cut} \approx \ell_{\rm max}/5$ gives good results.

\subsubsection{Time domain approaches}
\label{sec:GF-TD}

The frequency domain approach to computing the retarded Green function has several shortcomings. It
relies on relatively difficult technical calculations and has poor convergence properties at early
times and in scenarios where the worldline is not well-represented by a discrete spectrum of
frequencies. Recent work has shown that the Green function may also be accurately
computed (at least for the purposes of self-force calculations) in the time domain using
straightforward numerical evolutions. A time domain calculation sidesteps issues related to a wide
frequency spectrum and appears to exhibit much better convergence properties at early times.

There are two closely-related proposals for computing the Green function in the time domain.
In \cite{Zenginoglu:2012xe}, Eq.~\eqref{eq:BoxG-scalar} was numerically solved as an initial value problem,
with the delta-function source approximated by a narrow Gaussian. A reformulation of this approach
as a homogeneous problem with Gaussian initial data was subsequently given in \cite{Wardell:2014kea}. It
was found that these ``numerical Gaussian'' approaches are able to approximate the retarded Green
function remarkably well in a large region of the space required by worldline convolutions. The
size of the Gaussian limits the scale of the smallest features which can be resolved, but it turns
out that this is not significantly detrimental to a self-force calculation.

\begin{figure}[htb!]
\begin{center}
\includegraphics[width=8.5cm]{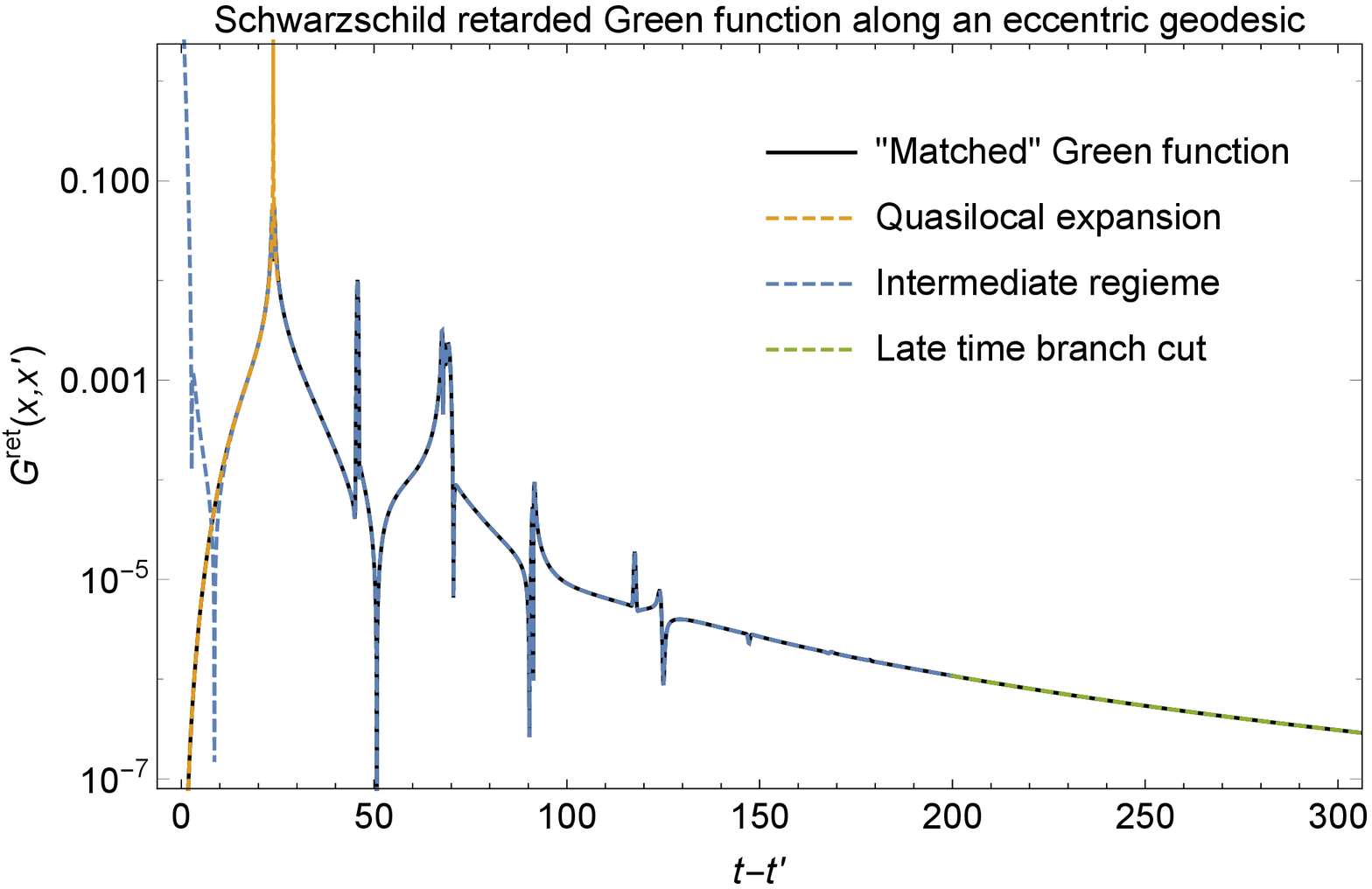}
\includegraphics[width=8.5cm]{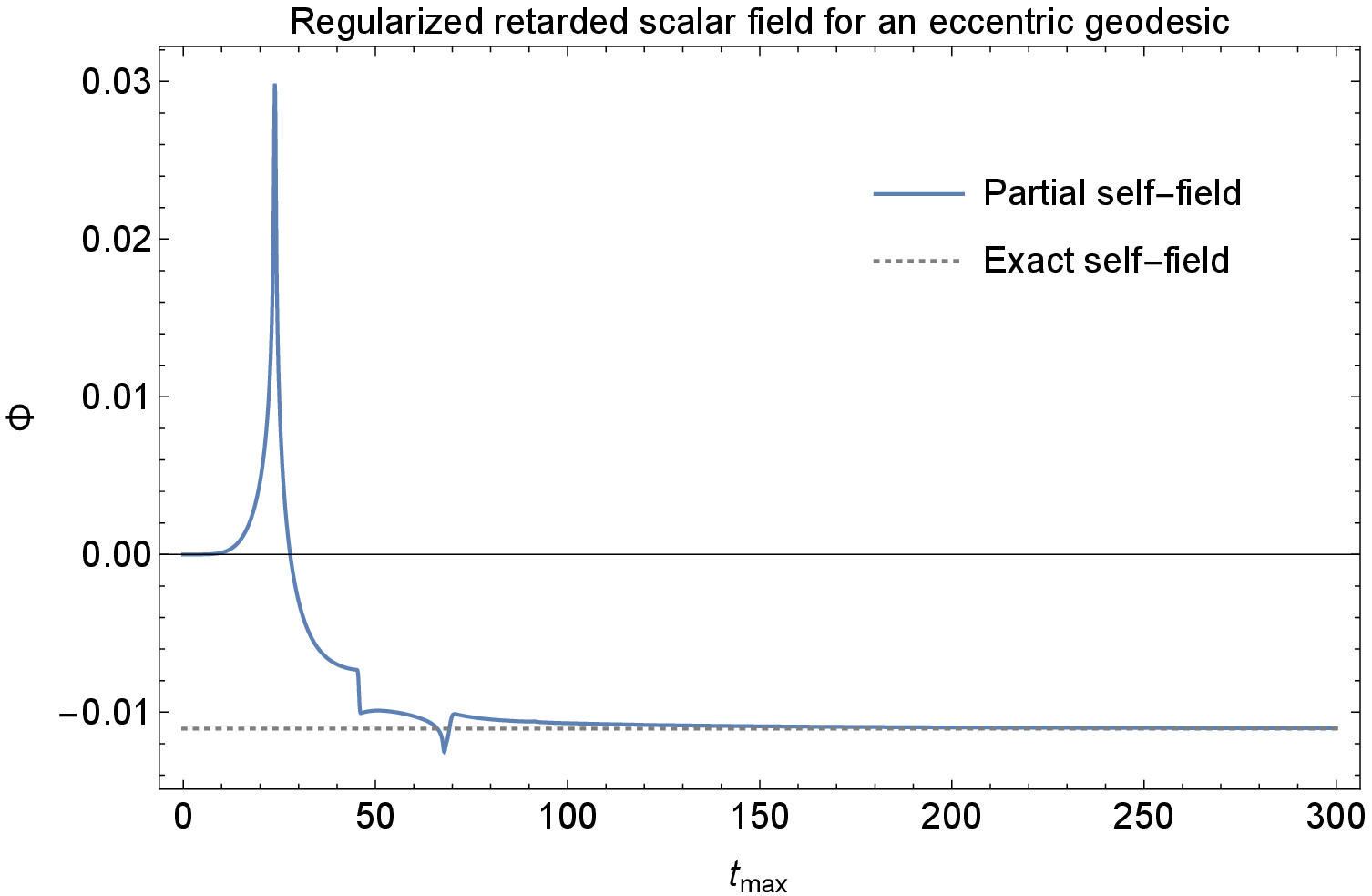}
\end{center}
\caption{
Top: Matched Green function along an eccentric orbit (eccentricity $e=0.5$ and semilatus rectum $p=7.2$) in Schwarzschild spacetime.
The full Green function (black) is constructed by combining a quasilocal approximation at early
times (orange, dashed line) with a time domain calculation at intermediate times (blue, dashed
line) and a late-time branch cut approximation at late times (green, dashed line).
Bottom: Integrating the retarded Green function (excluding the direct part) up to some time point on
the past worldline gives the contribution to the regularized self-field from all points on the past
worldline up to that time. With an eccentric orbit of period $\approx 317 M$, we see that a good
approximation to the self-field is obtained by including the contributions from less than one full
orbit. Note that the formal divergences of the Green function are integrable and do not cause
significant numerical difficulty in computing the self-force.
}
\label{fig:matched-GF}
\end{figure}

The only regimes where the numerical Gaussian approach is not well suited to computing the
retarded Green function are at very early and very late times. At early times the direct
$\delta(\sigma(x,x))$ term (which must be excluded from the worldline convolution of the retarded Green
function) is smeared out and is difficult to isolate from the rest of the Green function. There is
no conceptual difficulty at late times, but the reality of a numerical evolution is that it can
only be run for a finite time. Fortunately, both of these issues are easily overcome; the former by
using the quasilocal expansion at early times, and the latter by using a late-time expansion of the
branch cut integral (see Fig.~\ref{fig:matched-GF}).

In the time domain approach, each numerical time domain evolution gives the Green function
$G(x_0,x')$ for a single base point $x_0$, and for all source points, $x$. As a result, a single
numerical calculation can only be used to compute the self-force at a single spacetime point, but
it can be computed for any past-worldline ending up at that point. This is in contrast to other
self-force methods, where a single orbit is considered at a time, but a single calculation gives
the self-force at all points on the orbit. The problem of efficiently spanning the parameter space of base points $x_0$ is an ideal application of reduced order methods
\cite{Field:2013cfa,Wardell:Capra17}.

\subsection{Mode-sum regularization} \label{sec:msr}
The mode-sum regularization scheme, first proposed by Barack and Ori \cite{Barack:1999wf}, has
proven highly successful as a computational self-force strategy, having been applied to the
computation of the the self force for a variety of configurations in Schwarzschild
\cite{Barack:2000zq,Burko:2000xx,Detweiler:2002gi,DiazRivera:2004ik,Haas:2006ne,Haas:2007kz,
Canizares:2009ay,Canizares:2010yx,Jaramillo:2011gu,Barack:2007tm,Barack:2002ku,Sago:2008id,Sago:2009zz,Barack:2010tm,
Thornburg:2010tq,Keidl:2010pm,Shah:2010bi,Merlin:2014qda} and Kerr
\cite{Warburton:2010eq,Warburton:2011hp,Shah:2012gu,Warburton:2014bya,Isoyama:2014mja}
spacetimes. The success of the
method hinges on the fact that the first order retarded field diverges in a way which can be
effectively smoothed out by a spectral decomposition in the angular directions. More specifically,
the divergence appears as an odd power of $1/s$, where $s^2 = (g^{\alpha\beta} + u^\alpha u^\beta)
\sigma_{;\alpha} \sigma_{;\beta}$ is an appropriate measure of distance from the worldline. The
result is that the $1/s$ divergence of the field near the worldline turns into an infinite sum of
modes, each of which are individually finite (but possibly discontinuous) on the worldline. The
divergence of the field then manifests itself through the failure of the (infinite) sum over modes
to converge. Conveniently, this odd-in-$1/s$ property of the retarded field also holds for its
derivatives, both the first derivatives required for the self-force and higher derivatives which
are useful for computing higher-multipole gauge invariants \cite{Dolan:2013roa,Dolan:2014pja}. As a
result, an arbitrary number of derivatives of the first order retarded field is represented by an
infinite sum of modes, each of which are individually finite (but possibly discontinuous) on the
worldline. The trade-off is that as the number of derivatives increases the divergence of the sum
over modes becomes increasingly strong.

Since the individual modes are finite, numerical calculations of the retarded field and its
derivatives can be done in the reduced $(t,r)$ space without encountering any numerical
divergences. The remaining piece of the problem is a method for rendering the divergent sum over
modes finite. An analytic decomposition of the angular dependence of the Detweiler-Whiting singular
field yields ``regularization parameters'' which may be subtracted mode-by-mode from the numerical
retarded field values. Provided all parts of the Detweiler-Whiting singular field which don't
vanish on the worldline are included in the calculation, the regularized sum over modes is
convergent on the worldline and one never encounters any numerical divergences.

It is important to note that the use of the Detweiler-Whiting singular field is not merely a
convenience; it also provides well-motivated physical grounds to justify the mode-sum regularization
approach. Other ad-hoc approaches based on identifying the asymptotically divergent contributions
to the mode-sum often produce the correct regularization parameters, but their use is difficult to
justify rigorously and can easily lead to incorrect results. Crucially, there is no way of knowing
whether such an ad-hoc regularization procedure is producing a physically correct result, or if
important contributions are being overlooked. They should therefore not be relied on without
extreme care.

Unfortunately, despite its success in first order calculations, mode-sum regularization alone is
not an effective tool for computing the second order self-force. The reason for this is
straightforward: the second order retarded field contains divergent terms which appears in the form
of even powers of $1/s$. Intuitively, this arises from the fact that the second order field
contains terms which depend quadratically on the first order field. Unlike the odd-in-$1/s$ case,
the angular decomposition of $1/s^2$ leads to individual modes which diverge logarithmically as the
worldline is approached. Fortunately, all is not lost for the mode-sum method as a computational
tool at second order. Provided the leading order logarithmic divergence is subtracted by some other
means (for example, using the effective source approach), regularization parameters may be used to
accelerate the rate of convergence of the mode sum. Such a hybrid scheme complements the generality
of the effective source approach with the computational efficiency of mode-sum regularization.

\subsubsection{Regularization parameters}

\begin{figure}[htb!]
\begin{center}
\includegraphics[width=10cm]{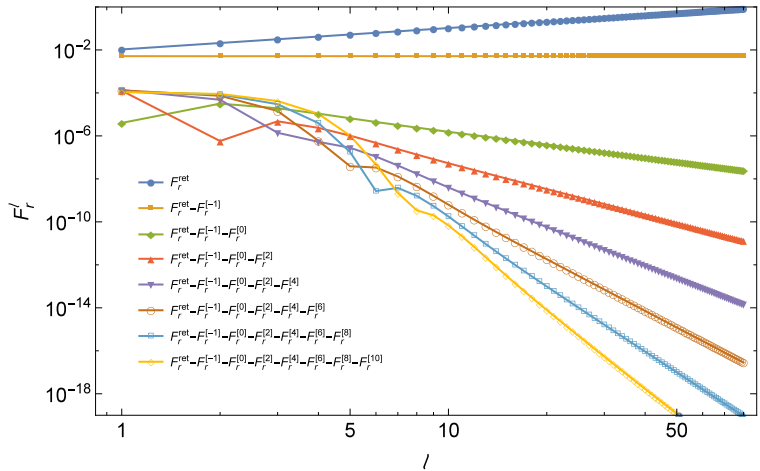}
\end{center}
\caption{
Spherical harmonic modes of the self-force for a point scalar charge on a
circular orbit of radius $r_0 = 6M$ in Schwarzschild spacetime. By subtracting
analytically determined high-order ``regularization parameters'', the sum over
modes is rendered more and more rapidly convergent. Each regularization parameter,
$F_r^{[n]}$, behaves asymptotically for large-$\ell$ as $1/\ell^{n}$.
}
\label{fig:modesum}
\end{figure}

The computation of regularization parameters has been addressed by a series of calculations
stemming from the original Barack-Ori derivation. Barack and Ori's original work gave the first two
self-force regularization parameters for scalar, electromagnetic and Lorenz-gauge gravitational
cases in both Schwarzschild and Kerr spacetimes. These are sufficient for computing the regularized
self-force, but they yield mode-sums which have relatively poor quadratic convergence with the
number of modes included. Since the Detweiler-Whiting regularized field is a smooth, homogeneous
function in the vicinity of the worldline, its mode-sum representation converges exponentially.
This spectral convergence is spoiled by the fact that only the portion of the Detweiler-Whiting
singular field which does not vanish on the worldline is subtracted by the leading order
regularization parameters. The mode decomposition effectively contains information about the
extension of the field off the worldline to the entire two-sphere, so the regularized field
contains residual pieces of the Detweiler-Whiting singular field off the worldline, but on a
two-sphere of the same radius. By deriving higher-order regularization parameters one can subtract
these residual pieces order-by-order, leaving a mode-sum which is more and more rapidly convergent
(see Fig.~\ref{fig:modesum}).
The derivation of these higher-order parameters is closely related to the computation of the
quasilocal expansion of the Green function and has been addressed in a series of papers: the first
higher order parameter was given in \cite{Detweiler:2002gi} for the case of a scalar charge on a circular
orbit in Schwarzschild spacetime, and for eccentric geodesic orbits in \cite{Haas:2006ne}. This
was subsequently extended by several further orders for equatorial geodesic motion in the scalar,
electromagnetic and gravitational cases in both Schwarzschild \cite{Heffernan:2012su} and Kerr
\cite{Heffernan:2012vj} spacetimes. Recent work has also produced parameters for accelerated
worldlines \cite{Casals:2012qq,Linz:2014pka}.

Within these derivations of high-order regularization parameters, a subtle ambiguity appears
through the elevation of the four-velocity from a quantity defined on a worldline to a quantity
defined everywhere on the two-sphere. A natural, covariant choice is to define this off-worldline
extension through parallel transport, $u^{\alpha'} = g_{\alpha}{}^{\alpha'} u^\alpha$. However, in
practical calculations it is often convenient to make a coordinate choice. For example, a common
choice is to define the extension in terms of ``constant coordinate components'', i.e. to define
$u^{\alpha'}$ such that its components in some coordinate system have a constant value everywhere on
the two-sphere. This is a perfectly valid choice, as is any other choice where $u^{\alpha'}$ agrees
with the actual four-velocity when evaluated on the worldline. The only caveat is that the
regularization parameters beyond the leading two orders change depending on the particular choice
of extension. As a result, in order to use higher-order regularization parameters it is essential
that a compatible choice of off-worldline extension is used in the retarded field calculation.

\subsubsection{Choice of basis and gauge}

There are two other factors which must be considered in the mode-sum scheme: the choice of angular basis functions for the spectral decomposition and the choice of gauge in the electromagnetic and
gravitational cases.

The choice of basis functions is typically motivated by their ability to produce separability of
the retarded field equations; an appropriate choice of basis functions yields an independent set
of equations in $(t,r)$ space for each individual mode. For the case of a scalar charge in Schwarzschild
spacetime, the appropriate basis functions are the standard spherical harmonics. In the Kerr case
the spheroidal harmonics are required for separability. For electromagnetic and gravitational cases,
there are several choices. For Schwarzschild spacetime one can choose between a tensor harmonic
basis and a basis of spin-weighted spherical harmonics. In the Kerr case only the spin-weighted
spheroidal harmonics are known to yield separability \cite{Teukolsky:1972my,Teukolsky:1973ha}.

This choice of angular basis also affects the regularization parameters. The parameters for scalar
spherical harmonic modes, tensor harmonic modes, spin-weighted spherical harmonic modes and
spheroidal harmonic modes are all potentially different. However, it is always possible project the
tensor, spin-weighted, or spheroidal harmonic modes onto the scalar spherical harmonic basis. In
fact, since the regularization parameters are most easily obtained for the scalar spherical
harmonic basis, calculations of the retarded field are typically done using the computationally
most convenient basis and the result is then projected onto the scalar spherical harmonics before
the regularization and sum-over-modes steps are done \cite{Barack:2010tm}.

A more difficult issue in the electromagnetic and gravitational cases is the
choice of gauge. The Detweiler-Whiting singular field is defined in Lorenz
gauge (since it is derived from a Lorenz gauge Green function), but numerical
calculations of the retarded field are more easily done in either radiation or
Regge-Wheeler gauge. The existence of tensor spherical harmonics makes a
Lorenz-gauge calculation possible, if somewhat cumbersome in the Schwarzschild
case \cite{Akcay:2010dx,Akcay:2013wfa,Osburn:2014hoa}. One obtains a coupled
set of 10 equations for the metric perturbation, but there is no coupling
between different different tensor-harmonic modes. Unfortunately, this does not
hold for the Kerr case as there are no known tensor spheroidal harmonics.
Rather than trying to derive tensor spheroidal harmonics, a better approach is
to work with the relatively straightforward Teukolsky equation in radiation
gauge \cite{Shah:2012gu}. The difficulty then is in identifying the appropriate regularization
parameters, particularly since the gauge transformation from Lorenz gauge is
itself often singular. This remained an open problem until recently, when an
understanding of how to apply the mode-sum regularization scheme in radiation
gauge was finally established \cite{Pound:2013faa} and applied in a self-force
calculation \cite{Merlin:2014qda}.

\subsubsection{Mode-sum regularization in the frequency domain}

The mode-sum scheme provides a method for computing the regularized self-force using numerical data
for the modes of the retarded field in the reduced $(t,r)$ space. There are, however, several
possibilities for computing this numerical data. One option relies on a further decomposition of
the time dependence into Fourier-frequency modes, in an analogous way to the frequency-domain Green
function described in Sec.~\ref{sec:GF-FD} above. Using a spin-weighted spheroidal harmonic basis,
this leads to a radial equation for the Teukolsky function,
\begin{equation}
  \Delta^{-s} \frac{d}{dr} \Big( \Delta^{s+1} \frac{dR}{dr}\Big) + V(r) R = \mathcal{T}(r)
\end{equation}
with the potential given by
\begin{equation}
  V(r) = \frac{K^2 - 2 i s (r - M) K}{\Delta} + 4 i r \omega s - \lambda.
\end{equation}
Here, $\Delta \equiv r^2 - 2 M r + a^2$, $K \equiv (r^2 + a^2) \omega - a m$, $\lambda$ is an
eigenvalue of the spheroidal equation,
\begin{align}
 \Big[\frac{1}{\sin \theta} \frac{d}{d\theta} \Big(\sin \theta \frac{d}{d\theta}\Big) &- a^2 \omega^2 \sin^2 \theta - \frac{(m+s\cos \theta)^2}{\sin^2 \theta} \nonumber \\
 & - 2 a \omega s \cos \theta + s + 2 m a \omega + \lambda] {}_{s} S_{\ell m} = 0
\end{align}
and $a$ and $M$ are the Kerr spin and mass parameters. For $a=0$, $\lambda = (\ell-s)(\ell+s+1)$
and this reduces to the Schwarzschild wave equation decomposed into spin-weighted spherical
harmonics (in radiation gauge for the gravitational case), while for $s=0$ it reduces to the scalar
wave equation. A decomposition of the point-particle source term yields a source involving
$\delta(r - r_p)$ (and in some cases its derivative), where $r_p$ is the radial location of the
particle. In practice, the frequency domain mode-sum approach proceeds in the same way as for the
Green function: two independent solutions of the homogeneous radial equation are obtained and are
matched at the particle's location\footnote{For eccentric orbits where the particle can not be
considered to be at a single radial location in the frequency domain this matching must be modified
slightly using, for example, the method of extended homogeneous solutions
\cite{Barack:2008ms,Hopper:2010uv}.}. Then, the distributional sources do not introduce any numerical
difficulty as they simply appear as jumps when matching the homogeneous inner and outer
solutions.

The solutions of the Teukolsky equation for a given $(\ell,m,\omega)$ can be obtained either
through straightforward numerical integration of the radial ordinary differential equation
(typically with some modifications to improve numerical accuracy \cite{Sasaki:1981kj,Sasaki:1981sx}) or as an
approximation in the form of an infinite convergent series of hypergeometric functions.
The latter method is based an idea originally developed by Leaver \cite{Leaver:1986JMP} and now commonly
referred to as the ``MST'' method, after Mano, Suzuki, and Takasugi \cite{Mano:1996vt} who reformulated
it into its current form. It provides an efficient and highly accurate method for computing
solutions of the radial equation. For example, recent results have used it to compute solutions
accurate to several hundred decimal places, allowing the solutions to be used to determine
previously unknown high-order post-Newtonian parameters \cite{Shah:2013uya,Shah:2014tka,Bini:2014nfa,Bini:2014ica,Bini:2014zxa}. For a
comprehensive review of the MST approach, see the living review by Sasaki and Tagoshi
\cite{Sasaki:2003xr} and references therein.

The frequency-domain approach is particularly appropriate in scenarios where the worldline is well
represented by a discrete spectrum involving a small number of frequencies. In such cases, mode-sum
regularization is by far the optimal choice and is unparalleled in its accuracy and computational
efficiency \cite{Barton:2008eb}. The prototypical example is a circular orbit, in which only a single frequency is
present. In the case of eccentric equatorial orbits (and inclined circular orbits in the Kerr
case), there are two fundamental frequencies, and also an infinite number of higher harmonics
produced from combinations of the fundamental frequencies. For mildly eccentric orbits this does
not cause a great deal of difficulty. However, for more eccentric orbits (with eccentricities $e
\gtrsim 0.5$) an increasingly large number of frequencies must be included and the competitive
advantage of the frequency domain approach is lost \cite{Barton:2008eb}. Even worse, generic geodesic orbits in Kerr
spacetime have three fundamental frequencies and the computational difficulty is so high that a
calculation has yet to be attempted. Similarly, unbound orbits and other cases such as radial infall are not well suited to frequency domain methods. Apart from these deficiencies, the frequency
domain mode-sum approach has been highly successful for producing results.

\subsubsection{Mode-sum regularization in the time domain}

The mode-sum scheme may also be applied in the time domain by skipping the Fourier decomposition
step and instead solving a set of $1+1$D partial differential equations in $(t,r)$ space. The main
difficulty then is in appropriately handling the distributional source term which has the form
$\delta(r - r_p(t))$. One solution, used in
\cite{Barack:2007tm,Barack:2010tm,Haas:2007kz,Haas:2011np,Sundararajan:2007jg,Sundararajan:2008zm}, is to use a discretised
representation of a delta function and to construct the computational grid such that the worldline
only ever passes \emph{between} grid points.

An alternative approach is to reformulate the problem in an
analogous way to the frequency-domain method. By splitting the computational domain into two
domains --- one either side of the particle --- the delta-function source can be reformulated in
terms of a jump in the fields and their derivatives at the interface of the two domains. This
method is well suited to highly-accurate spectral numerical methods as all of the numerically
evolved fields are smooth functions. It was implemented using discontinuous-Galerkin methods in
\cite{Field:2009kk,Field:2010xn}
and using Chebyshev pseudo-spectral methods in \cite{Canizares:2009ay,Canizares:2010yx}. Eccentric orbits present a small
additional complexity in this case as the particle must always lie on the domain interface. This is
easily achieved by introducing a time-dependent mapping between the computational and physical
coordinates of the system \cite{Canizares:2010yx,Field:2010xn}.

\subsubsection{Limitations of the mode-sum regularization scheme}

Despite its resounding success to date, the mode-sum regularization scheme has some unfortunate
disadvantages which make it ill-suited as a general-purpose method for self-force calculations:
\begin{enumerate}
\item Its application in the Kerr case is only straightforward in the frequency domain, since the
  field equations are not separable in the time domain\footnote{It may still be possible to use
  mode-sum regularization for the Kerr case in the time domain by decomposing the field equations
  into \emph{spherical} harmonics and evolving the resulting infinitely coupled set of $1+1$D
  partial differential equations in a similar manner to the Schwarzschild case.}
\item It relies on the use of Lorenz gauge for regularization in the gravitational case. This is
  not a major issue in the Schwarzschild case since the tensor spherical harmonics may be used to
  decouple the Lorenz gauge field equations in the angular directions (leaving 10 coupled $1+1$D
  equations for each $\ell,m$ mode). However, there are no know tensor spheroidal harmonics which
  would be required for the Kerr case, and even if there were they would likely not be applicable
  in the time domain (again, it is conceivable that a coupled system of equations involving tensor
  spherical harmonics could be used, but the coupling would result in considerable complexity).
\item It is not applicable beyond first perturbative order, since the modes of the second
  order perturbation diverge logarithmically near the worldline.
\end{enumerate}
The first two issues are not necessarily showstoppers. There have been several attempts at $1+1$D
time-domain implementations using coupled spherical-harmonic modes in the Kerr case
\cite{Dolan:2012yt,Stein:Thesis}, and recent progress on reformulating mode-sum regularization for radiation gauge
has clarified the regularization issue \cite{Pound:2013faa}. The third point, however, appears insurmountable. For these
reasons, among others, the effective source method, described in the next section, was developed.

\subsection{Effective source approach}

Proposed in 2007 as a solution to the shortcomings of mode-sum regularization, the effective source
approach\footnote{Note that the effective source proposed by Lousto and Nakano
\cite{Lousto:2008mb} is similar in spirit, but differs in that it is not derived from the
Detweiler-Whiting singular field.} provides an alternative method for handling the divergence of
the retarded field. Rather than first computing the retarded field and then subtracting the
singular piece as a post-processing step, one can instead work directly with an equation for the
regular field. This idea --- independently proposed by Barack and Golbourn \cite{Barack:2007jh} and
by Vega and Detweiler \cite{Vega:2007mc} --- has the distinct advantage of involving only regular
quantities from the outset, making it applicable in a wider variety of scenarios than the mode-sum
scheme. In particular, since it does not rely on a mode decomposition of the retarded field, it can
be used by any numerical prescription for solving the retarded field equations, whether in the
frequency domain (where the method is really just a generalisation of mode-sum regularization) or
in the time domain as a $1+1$D, $2+1$D or even $3+1$D problem.

\begin{figure}[htb!]
\begin{center}
\includegraphics[width=6cm]{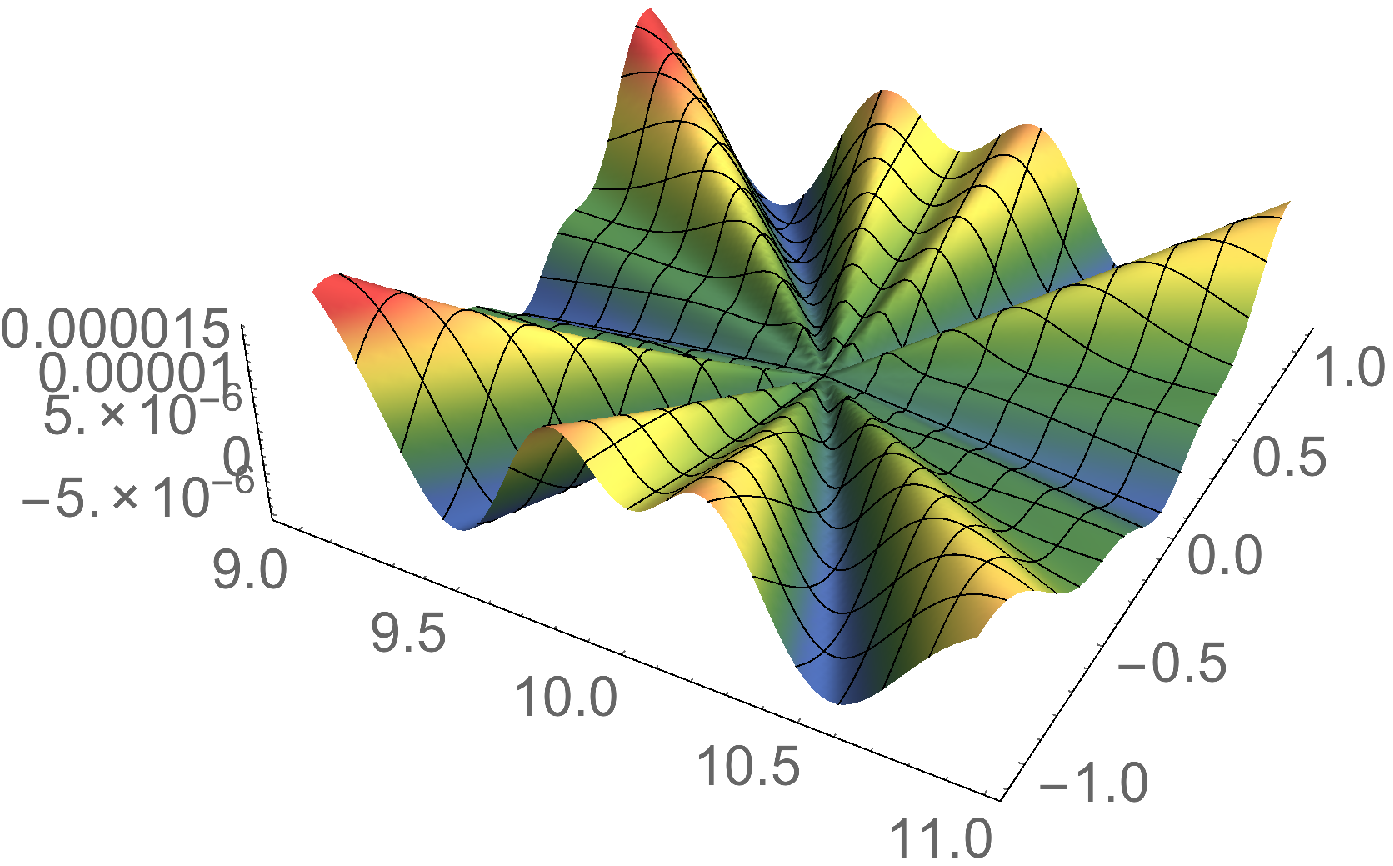}
\includegraphics[width=6cm]{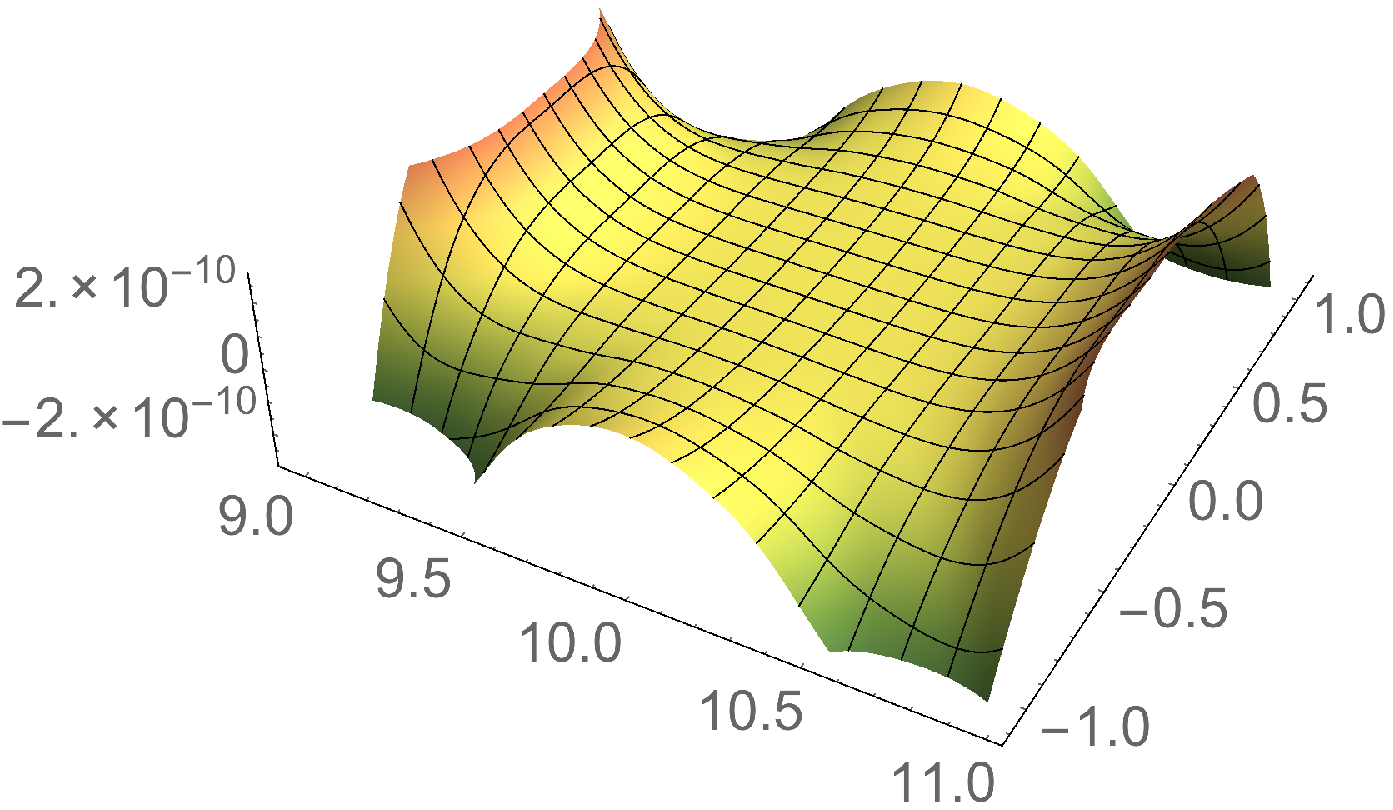}
\end{center}
\caption{
The effective source for a scalar particle on a circular orbit in Schwarzschild spacetime. The
source is generically non-smooth in the vicinity of the worldline (left), but the smoothness can be
improved by incorporating higher-order parts of the Detweiler-Whiting singular field into the
source.
}
\label{fig:effsource}
\end{figure}

\begin{figure}[htb!]
\begin{center}
\includegraphics[width=8cm]{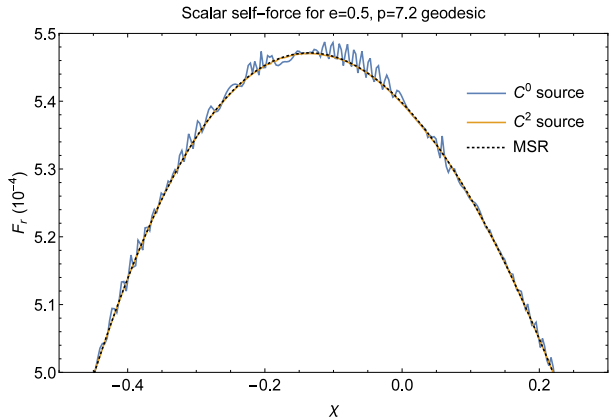}
\end{center}
\caption{
Radial self-force for a scalar particle on an eccentric orbit in Schwarzschild spacetime, computed
with a $3+1$D implementation of the effective source scheme. Here, the independent variable $\chi$ is a
``relativistic anomaly parameter'' defined through $r = p M / (1+e \cos \chi)$, with $p$ the
semilatus rectum and $e$ the eccentricity.
The high-frequency errors using a continuous
but non-differentiable source (blue) are dramatically decreased by using a twice differentiable source (orange)
obtained from a higher-order approximation to the
Detweiler-Whiting singular field. For reference, a highly accurate value computed using frequency-domain
mode-sum regularization is also included (dashed, black). Figure based on version presented in Ref.~\cite{Diener:APS}.
}
\label{fig:src-smoothness}
\end{figure}

The basic idea is to use the split of the retarded field into regular and singular pieces,
Eq.~\eqref{eq:singular-regular-split}, to rewrite the field equations, Eqs.~\eqref{eq:field-gravity},
\eqref{eq:field-scalar} and \eqref{eq:field-em} as equations for the Detweiler-Whiting regular
field,
\begin{subequations}
  \label{es:effsource-field}
\begin{align}
  (\Box&-\xi R)\Phi^{\rm R} = (\Box-\xi R)(\Phi^{\rm ret} - \Phi^{\rm S})  \nonumber \\
                &= -4\pi\rho - (\Box-\xi R) \Phi^{\rm S},
\end{align}
\begin{align}
  (\Box \delta_\alpha{}^\beta &- R_\alpha^\beta) A^{\rm R}_\beta =
  (\Box \delta_\alpha{}^\beta - R_\alpha^\beta) (A^{\rm ret}_\beta-A^{\rm S}_\beta) \nonumber \\
    &= - 4 \pi e \int g_{\alpha\alpha'} u^{\alpha'} \delta_4(x, z(\tau')) d \tau' - (\Box \delta_\alpha{}^\beta - R_\alpha^\beta) (A^{\rm S}_\beta),
\end{align}
\begin{align}
   (\Box & \delta_\alpha{}^\gamma \delta_\beta{}^\delta + 2 C_{\alpha}{}^{\gamma}{}_{\beta}{}^{\delta }) \bar{h}^{\rm R}_{\gamma\delta} =
   (\Box \delta_\alpha{}^\gamma \delta_\beta{}^\delta + 2 C_{\alpha}{}^{\gamma}{}_{\beta}{}^{\delta }) (\bar{h}^{\rm ret}_{\gamma\delta} - \bar{h}^{\rm S}_{\gamma\delta}) \nonumber \\
   &= - 16 \pi \mu \int g_{\alpha'(\alpha} u^{\alpha'} g_{\beta)\beta'} u^{\beta'} \delta_4(x, z(\tau')) d \tau' - (\Box \delta_\alpha{}^\gamma \delta_\beta{}^\delta + 2 C_{\alpha}{}^{\gamma}{}_{\beta}{}^{\delta }) (\bar{h}^{\rm S}_{\gamma\delta}).
\end{align}
\end{subequations}
If the singular field used in the subtraction is exactly the Detweiler-Whiting singular field, then
the two terms on the right hand side of this equation cancel and the regularized field would be a
homogeneous solution of the wave equation. Unfortunately, one typically does not have access to an
exact expression for the singular field. Indeed, the Detweiler-Whiting singular field is only
defined through a Hadamard parametrix which is not even defined globally. Instead, the best one can
typically do is a local expansion which is valid only in the vicinity of the worldline. Borrowing
the language of Barack and Golbourn, we refer to an approximation to the singular
field as a ``puncture'' field, $\Phi^{\rm S} \approx \Phi^{\rm P}$, $A_\alpha^{\rm S}
\approx A_\alpha^{\rm P}$, $\bar{h}_{\alpha\beta}^{\rm S} \approx \bar{h}_{\alpha\beta}^{\rm P}$.
Then, the corresponding approximate regular field --- referred to as the ``residual field'' --- is
no longer a solution of the homogeneous equation, but instead is a solution of the sourced equation
with an \emph{effective source} (see Fig.~\ref{fig:effsource}) which is defined to be the right-hand
side of Eq.~\eqref{es:effsource-field} with the puncture field substituted for the singular field
\begin{subequations}
\begin{align}
  S^{\rm eff} &= - 4 \pi q \int \delta_4(x, z(\tau')) d \tau' - (\Box-\xi R)\Phi^P, \\
  S_\alpha^{\rm eff} &= - 4 \pi e \int g_{\alpha\alpha'} u^{\alpha'} \delta_4(x, z(\tau')) d \tau' - (\Box \delta_\alpha{}^\beta - R_\alpha^\beta) (A^{\rm P}_\beta), \\
  S_{\alpha\beta}^{\rm eff} &= - 16 \pi \mu \int g_{\alpha'(\alpha} u^{\alpha'} g_{\beta)\beta'} u^{\beta'} \delta_4(x, z(\tau')) d \tau' - (\Box \delta_\alpha{}^\gamma \delta_\beta{}^\delta + 2 C_{\alpha}{}^{\gamma}{}_{\beta}{}^{\delta }) (\bar{h}^{\rm P}_{\gamma\delta}).
\end{align}
\end{subequations}
Note that the presence of a distributional component of the source on the worldline is merely a
formal prescription; in practice the puncture field is chosen so that it exactly cancels this
distributional component on the worldline. This effective source is then finite everywhere, but has
limited differentiability on the worldline. This makes it well-suited to numerical implementations
since no divergent quantities are ever encountered. The only numerical difficulty arises from the
non-smoothness of the source in the vicinity of the worldline (see Fig.~\ref{fig:effsource}), which
leads to numerical noise in the computed self force. The noise can be reduced by making the source
smoother using higher-order parts of the Detweiler-Whiting singular field. As shown in
Fig.~\ref{fig:src-smoothness}, at the same numerical resolution a higher order source ($C^2$ in
this case) eliminates the vast majority of the numerical noise that is present when using a lower
order source ($C^0$ in the case in the figure). The cost of this improved accuracy is a source which
is considerably more complicated, and costly to compute in a numerical code.

An additional level of complexity arises from the fact that the puncture field is defined only in
the vicinity of the worldline. To avoid ambiguities in its definition far from the worldline, one
must ensure that the puncture field goes to zero there. This is most easily achieved by multiplying
it by a window function, $\mathcal{W}$, with properties such that it only modifies the puncture
field in a way that its local expansion about the worldline is preserved to some chosen order.
In a first-order calculation of the self-force, it suffices to choose $\mathcal{W}$ such that
$\mathcal{W}(x_p) = 1$, $\mathcal{W'}(x_p) = 0$, $\mathcal{W''}(x_p) = 0$ and $\mathcal{W} = 0$ far
away from the worldline.
\begin{figure}[htb!]
\begin{center}
\includegraphics[width=8cm]{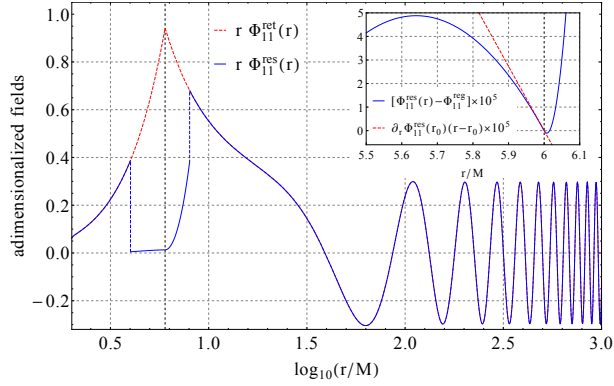}
\end{center}
\caption{
The $\ell=1, m=1$ spherical-harmonic mode of the residual field, $\Phi^{\rm res}$, for a scalar point particle on a circular orbit in Schwarzschild spacetime. This was produced in
\cite{Warburton:2013lea} using a frequency-domain effective source approach.
}
\label{fig:residual}
\end{figure}
The residual field (see Fig.~\ref{fig:residual}) then obeys
\begin{subequations}
\begin{align}
  (\Box-\xi R)\Phi^{\rm res} &= S^{\rm eff} \label{eq:field-residual-scalar} \\
  (\Box \delta_\alpha{}^\beta - R_\alpha^\beta) A^{\rm res}_\beta &= S_{\alpha}^{\rm eff} \label{eq:field-residual-em} \\
  (\Box \delta_\alpha{}^\gamma \delta_\beta{}^\delta + 2 C_{\alpha}{}^{\gamma}{}_{\beta}{}^{\delta }) \bar{h}^{\rm res}_{\gamma\delta} &= S_{\alpha\beta}^{\rm eff} \label{eq:field-residual-gravity}
\end{align}
\end{subequations}
and has the properties
\begin{subequations}
\begin{gather}
\Phi^{\rm res} (x_p) = \Phi^{\rm R}(x_p), \quad \nabla_\alpha \Phi^{\rm res}(x_p)   = \nabla_\alpha \Phi^{\rm R}(x_p),
\nonumber \\
  \Phi^{\rm res}(x) = \Phi^{\rm ret}(x) \quad \text{for} \quad x \not\in \operatorname{supp}(\mathcal{W}),
\end{gather}
\begin{gather}
A_{\alpha}^{\rm res} (x_p) = A_{\alpha}^{\rm R}(x_p), \quad \nabla_\alpha A_{\alpha}^{\rm res}(x_p)   = \nabla_\alpha A_{\alpha}^{\rm R}(x_p),
\nonumber \\
  A_{\alpha}^{\rm res}(x) = A_{\alpha}^{\rm ret}(x) \quad \text{for} \quad x \not\in \operatorname{supp}(\mathcal{W}),
\end{gather}
\begin{gather}
h_{\alpha\beta}^{\rm res} (x_p) = h_{\alpha\beta}^{\rm R}(x_p), \quad \nabla_\alpha h_{\alpha\beta}^{\rm res}(x_p)   = \nabla_\alpha h_{\alpha\beta}^{\rm R}(x_p),
\nonumber \\
  h_{\alpha\beta}^{\rm res}(x) = h_{\alpha\beta}^{\rm ret}(x) \quad \text{for} \quad x \not\in \operatorname{supp}(\mathcal{W}).
\end{gather}
\end{subequations}

As the residual field coincides with the retarded field far from the particle we can use the usual
retarded field boundary conditions when solving Eqs.~\eqref{eq:field-residual-scalar}, \eqref{eq:field-residual-em} and \eqref{eq:field-residual-gravity}. The details of a
numerical implementation of the effective source approach are then much the same as for the Green
function and mode-sum regularization schemes. One can use either a frequency domain or a time domain
method for solving the field equations, the key differences now being that there is no restriction
to $1+1$D, and that the effective source must be included as a source term. A more
thorough review of the effective source approach --- including a detailed description of methods for obtaining explicit expressions for the effective source --- can be found in Refs.~\cite{Wardell:2011gb,Vega:2011wf}.

The effective source approach has been successfully applied in the frequency domain
\cite{Warburton:2013lea}, and in the time domain in $1+1$D \cite{Vega:2007mc}, $2+1$D
\cite{Dolan:2010mt,Dolan:2011dx,Dolan:2012jg} and $3+1$D
\cite{Vega:2009qb,Diener:2011cc,Vega:2013wxa} contexts. It has also been formulated --- but not yet
implemented --- in the gravitational case to second order in the mass ratio \cite{Pound:2014xva,Gralla:2012db};
for the second order problem it is currently the only viable computational strategy. Since the
effects of the second-order metric perturbation will be very small --- being suppressed by two orders
of the mass ratio relative to test body effects --- it is likely that a highly accurate numerical
scheme will be required, suggesting a frequency domain treatment of the problem where one
encounters \emph{ordinary differential equations} (ODEs) which are relatively easy to solve
numerically to high accuracy.

\section{Evolution schemes}
\label{sec:evolution}

The calculation of the self-force is only the first stage in the production of an inspiral model.
Another critical component is a scheme for evolving the orbit using this self-force information.
Various approaches have been proposed, each of which brings with it its own advantages and
disadvantages. Approaches which make approximations in the self-force used to drive the inspiral
can give substantial decreases in the computational cost of an inspiral simulation, but come at the
cost of ignoring potentially relevant physical effects.

\subsection{Dissipation driven inspirals}

A straightforward model for the inspiral can be obtained from energy balance considerations. Using
a flux calculation --- in which fluxes of energy and angular momentum are obtained by evaluating
the point-particle retarded field near the horizon and out at infinity -- the entire issue of
regularization is avoided and one obtains an approximation to the contributions to the inspiral
coming from dissipative self-force effects. However, this is inadequate for capturing all relevant
effects from the first-order self-force. Being a dissipative approximation, it completely misses all
conservative corrections to the motion. Furthermore, there is no well-defined way of associating
fluxes far from the worldline with an instantaneous local self-force on the worldline; a flux model
can only be used to drive an inspiral in a time-averaged sense, ignoring some potentially important
dissipative contributions. Nevertheless, the relative straightforwardness and computational
efficiency of their implementation have made flux models a compelling approach to assessing
qualitative features of self-force driven inspirals. These ``kludge'' models have been used to
produce kludge waveforms for EMRI systems which capture at least some of the
qualitative physical effects
\cite{Glampedakis:2002ya,Hughes:2005qb,Drasco:2005kz,Sundararajan:2007jg,Sundararajan:2008zm}.

\begin{figure}[htb!]
\begin{center}
\includegraphics[width=6cm]{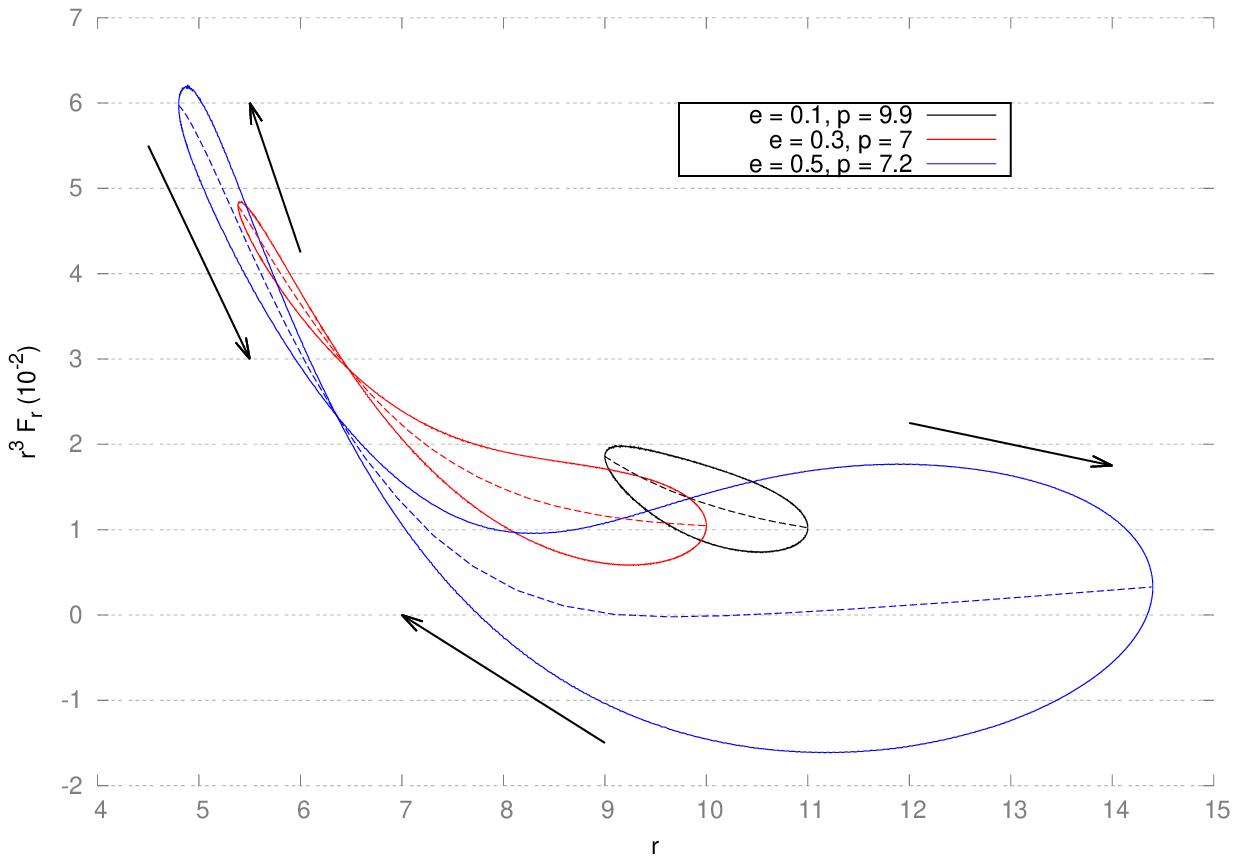}
\includegraphics[width=6cm]{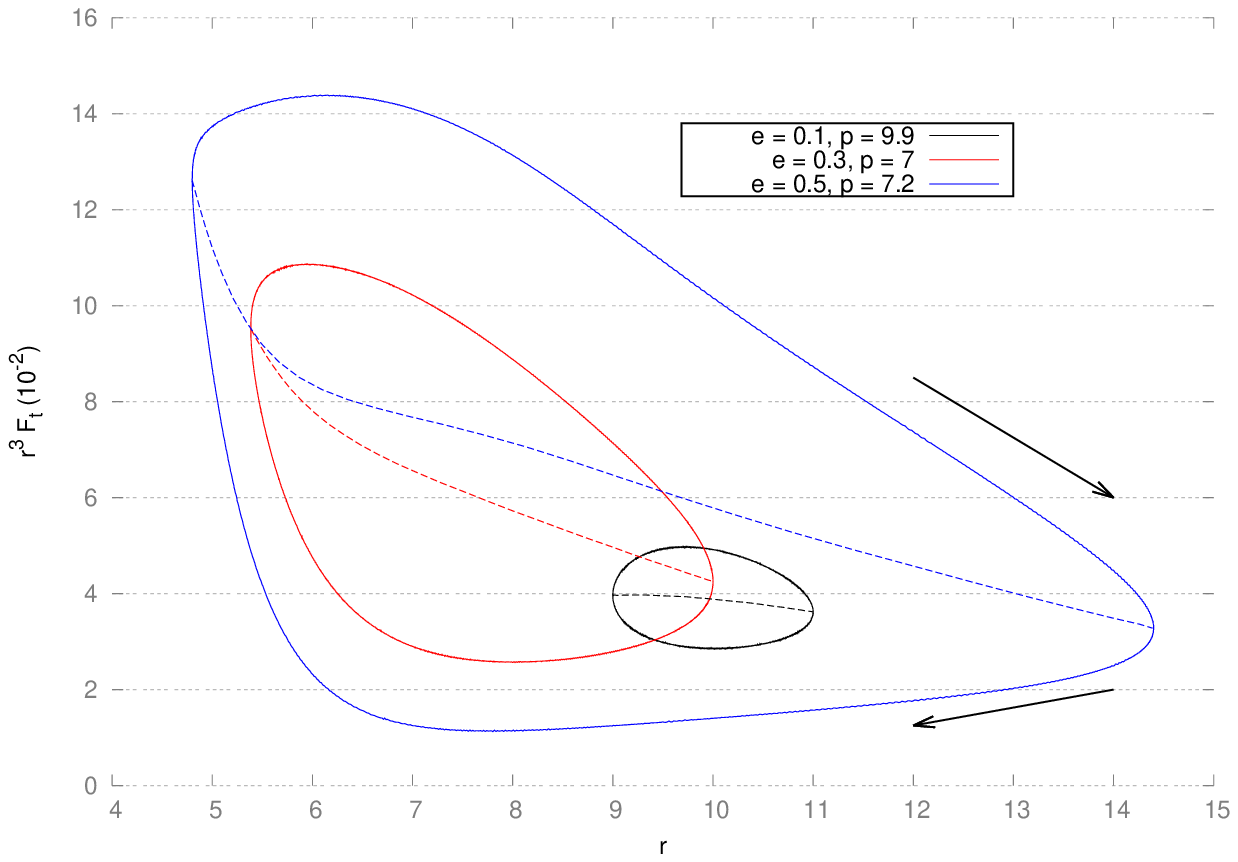}
\end{center}
\caption{
Radial (left) and time (right) components of the self-force through one radial
cycle, for three different geodesic orbits in Schwarzschild spacetime. Solid lines
indicate the full self-force and dashed lines indicate the conservative-only (left)
or dissipative-only (right) pieces. Arrows denote the direction of the geodesic
motion. Reproduced from Ref. \cite{Vega:2013wxa}.
}
\label{fig:loops}
\end{figure}

To improve on the flux-based dissipative model, one can instead make use of a
half-retarded minus half-advanced scheme \cite{Mino:2003yg,Gralla:2005et,Hinderer:2008dm,Flanagan:2012kg} to
compute a local, instantaneous dissipative self-force (see Fig.~\ref{fig:loops}).
This approach captures all dissipative effects responsible for driving the
inspiral, but still neglects small but potentially important
conservative corrections to the orbital phase of the system \cite{Warburton:2011fk}.

\subsection{Osculating self-forced geodesics}

In order to account for both conservative and dissipative effects from the first order self-force, it is essential to build an
inspiral model around a full calculation of the local self-force. This, however, still leaves some
flexibility in the choice of method for computing this local self-force. One option, based on the
osculating geodesics framework \cite{Pound:2007th} is particularly compelling as it allows for dramatic
improvements in computational efficiency by separating the coupled problem of simultaneously
solving for the orbit and for the regularized retarded field into two independent, largely
uncoupled computational problems. The basic idea is to perturbatively expand the worldline about a 
geodesic of the background spacetime, $z(\tau) = z_0(\tau) + \cdots$. Then, the self-force at first order is a
functional not of the evolving orbit $z(\tau)$, but of the geodesic orbit, $z_0(\tau)$, which is
instantaneously tangent to the worldline. This expansion is valid in the adiabatic regime, where
the orbit is evolving slowly and the difference between the geodesic and evolving orbits is small;
the error introduced by the approximation appears at the same order in the equations of motion
as a second-order perturbative correction to the field equations.

The improvements in computational efficiency brought about by the osculating geodesics
approximation are dramatic. The problem of self-consistently computing the regularized self-force
coupled to an arbitrarily evolving orbit is reduced to the much simpler problem of determining the
self-force for geodesic worldlines. Since orbits of black hole spacetimes are parametrised by at
most three conserved quantities (energy and angular momentum in the Schwarzschild case,
supplemented by the Carter constant in the Kerr case), it is computationally tractable to span the
entire parameter space of moderately-eccentric geodesic orbits. Even better, the
highly-accurate frequency domain mode-sum method can be used because bound geodesic orbits are
efficiently represented by a small frequency spectrum. In Ref.~\cite{Warburton:2011fk} this approach was explored
over an entire radiation-reaction timescale for moderately-eccentric inspirals in the Schwarzschild
spacetime\footnote{See also Ref.~\cite{Lackeos:2012de} for an approximate version which assumes a sequence of
quasi-circular orbits.}. This was achieved by tabulating the values for the self-force in a relevant portion of
the energy-angular momentum phase space of geodesic orbits, and using an interpolated model of the
tabulated results to drive an orbital inspiral.

\subsection{Self-consistent evolution}

Unfortunately, the adiabatic approximation responsible for the dramatic improvements in
computational efficiency brought by the osculating geodesics framework also introduces errors in the equation of motion at
second perturbative order, making the method inadequate for the purposes of precision EMRI astrophysics.
One possible solution, implemented in \cite{Diener:2011cc} for the
scalar field case, is to avoid expanding out the worldline and instead evolve the self-consistent
coupled system of equations, Eqs.~\eqref{eq:field-gravity} and \eqref{eq:accel-gravity};
\eqref{eq:field-scalar}, \eqref{eq:accel-scalar} and \eqref{eq:mdot-scalar}; or
\eqref{eq:field-em} and \eqref{eq:accel-em}. This is a computationally much more difficult problem as the field
equations must be solved in the time domain and one cannot rely on an efficient off-line tabulation
of self-force values. This self-consistent evolution can in principal be implemented using a
time-domain mode-sum scheme, but in practice implementations have instead used the effective source
approach since that will be necessary at second perturbative order (see Sec.~\ref{sec:msr} for an explanation of this issue).

While a self-consistent evolution incorporates all effects contributing to the first order self-force, and only
neglects second order contributions from the second order field equations, this comes at the cost
of computational efficiency. Whereas an osculating geodesics framework can evolve a large number
($\sim 10,000$) of orbits with ease once the initial off-line tabulation phase is complete
\cite{Warburton:2011fk}, the
self-consistent scheme requires a long calculation of the solutions of the first-order field
equations for each new orbit. In reality, this limits the practicality of the scheme to tens of
orbits using current methods. The self-consistent evolution scheme is therefore most valuable as
a benchmark against which other, less accurate but more efficient methods can be validated. In fact,
comparisons between the self-consistent and osculating geodesics scheme for the scalar case indicate
that the osculating geodesics scheme performs remarkably well \cite{Warburton:Capra17}.

\subsection{Two-timescale expansions}

The osculating geodesics scheme is fast, but inaccurate, while the self-consistent evolution scheme
is accurate, but slow. This begs the question of whether there is a middle ground which
incorporates most of the accuracy of a self-consistent evolution while maintaining the
computational efficiency of the osculating geodesics scheme. One promising possibility is that the
use of a two-timescale expansion could give just such a scheme. In the two-timescale scheme, rather
than using an expansion about a background geodesic (which is valid over short timescales
characterised by the orbital motion), one introduces an additional radiation reaction timescale
into the problem and incorporates the relevant effects over this radiation reaction timescale.

The relevant two-timescale expansion of the worldline equations of motion was completed in
Ref.~\cite{Hinderer:2008dm}, and follow-up work has improved our understanding of
important resonance effects for inspiral orbits
\cite{Flanagan:2010cd,Hirata:2010xn,Grossman:2011ps,Grossman:2011im,Gair:2011mr,Flanagan:2012kg,Isoyama:2013yor,Brink:2013nna,Ruangsri:2013hra,vandeMeent:2013sza,vandeMeent:2014raa}.
In order to consistently incorporate a two-timescale expansion into an orbital evolution scheme,
it will also be necessary to have a two-time expansion of the field equations. Promising progress
towards such an expansion was recently reported in \cite{Moxon:Capra17}, indicating that it is likely
that a self-consistent orbital evolution using a two-timescale expansion is indeed feasible.

\section{Discussion}

The techniques described in this review represent the current state of the art of self-force
calculations. The three primary approaches: worldline convolutions, mode-sum regularization and the
effective source approach can be considered complimentary, with each having regimes where they are
most appropriate:
\begin{itemize}
  \item The mode-sum scheme gives unparalleled accuracy (particularly in the frequency domain) for
  orbits which have a small frequency spectrum.
  \item The worldline convolution method provides valuable physical insight and can be easily
  applied to arbitrary orbital configurations, including those inaccessible by other means.
  \item The effective source approach can be used in arbitrary spacetimes without relying on any
  symmetries, and also stands out as the most applicable to a second order calculation.
\end{itemize}
There still remain important developments to be made in each case. For example, the mode sum and
effective source schemes
are still under development for the Kerr gravitational case, and worldline convolution
approaches have yet to be fully applied to the gravitational case for any spacetime.

Finally, it should be pointed out that this review is not an exhaustive exposition of all self-force
computation strategies. Many other calculations have not been described, including alternative
regularization strategies \cite{Rosenthal:2003qr,Rosenthal:2004wp,Kol:2013tfa}, 
near-horizon waveform calculations \cite{d'Ambrosi:2014iga,Hadar:2009ip,Hadar:2011vj,Hadar:2014dpa}, methods
based on effective field theory \cite{Birnholtz:2013nta,Birnholtz:2013ffa,Birnholtz:2014fwa,Birnholtz:2014gna}, analytic
calculations in particularly simple cases \cite{Wiseman:2000rm,Cho:2007jj,Ottewill:2012aj}, black holes in higher
dimensions \cite{Beach:2014aba}, and calculations in
non-black hole spacetimes such as wormholes \cite{Taylor:2012mv,Taylor:2014vsa} and cosmological
models \cite{Burko:2002ge,Haas:2004kw}. More details can be found in the reviews
\cite{Poisson:2011nh,Detweiler:2005kq}, and references therein.

\section*{Acknowledgements}
I am grateful to Scott Field and David Nichols for helpful discussions during the preparation of
this article, to Peter Diener for providing the data used in Fig.~\ref{fig:src-smoothness},
and to Niels Warburton, Scott Field and Eanna Flanagan for helpful comments on an early draft of this paper.

This material is based upon work supported by the National Science Foundation
under Grant Number 1417132. B.W. was supported by Science Foundation Ireland
under Grant No. 10/RFP/PHY2847, by the John Templeton Foundation New Frontiers
Program under Grant No. 37426 (University of Chicago) - FP050136-B (Cornell
University), and by the Irish Research Council, which is funded under the
National Development Plan for Ireland.

\bibliographystyle{unsrt}
\bibliography{barry_wardell_eom_proceedings_2013}

\end{document}